Proceedings of

# IX International Workshop on
# Locational Analysis and
# Related Problems

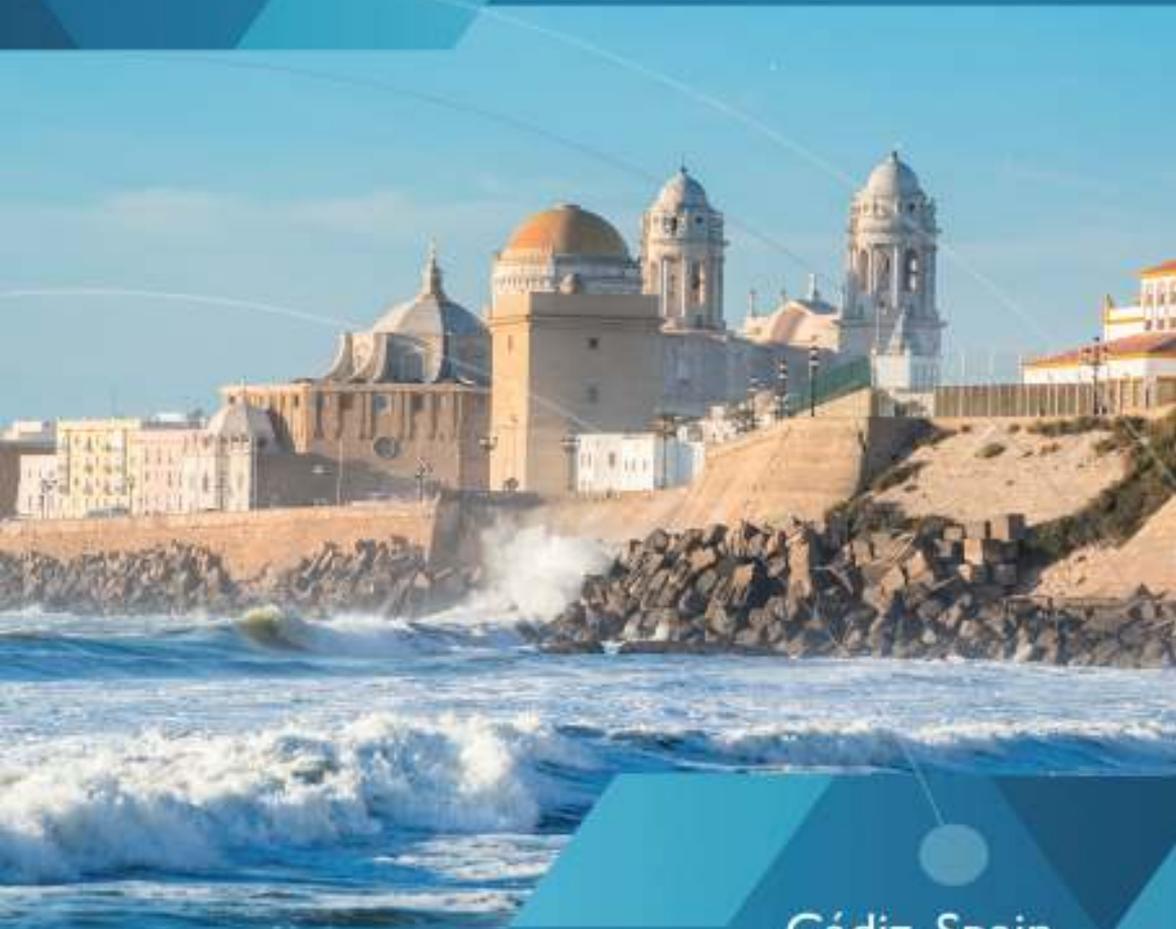

Cádiz, Spain
January 30-February 1, 2019

| | Wednesday Jan 30th | Thursday Jan 31st | Friday Feb 1st |
|---|---|---|---|
| 9:00 | | Session 3: Continuous Location | Session 9: Applications/ Routing/ Hub Location |
| 10:40 | | Coffee break | Coffee break |
| 11:10 | | Invited Speaker: Francisco Saldanha da Gama | Invited Speaker: Ivana Ljubic |
| 12:30 | | Session 4: | Session 10: Bilevel Location |
| 13:50 | | Routing | Location Network Meeting |
| 14:10 | | Lunch | |
| 14:30 | | | Lunch |
| 15:30 | | Session 5: | |
| 16:00 | Registration | Networks Design | |
| 16:30 | Opening | Break | |
| 16:40 | Session | Session 6: | |
| 16:45 | Session 1: | Networks Design II | |
| 17:40 | Discrete | Coffee break | |
| 18:00 | Location | Session 7: | |
| 18:25 | Coffee break | Applications | |
| 18:45 | Session 2: | | |
| 19:00 | Networks | Break | |
| 19:10 | | Session 8: | |
| 20:05 | | Discrete Location II | |
| 20:10 | | | |
| 21:30 | Welcome Reception | Dinner | |

# Conference Map

## Conference Venue

Restaurante Arsenio Manila -
Conference Dinner

Restaurante Arteserrano -
Lunch

La Calle del Libre Albedrío -
Welcome Reception

Hotel Spa Cádiz Plaza
- Spanish Location Network
Hotel

Facultad de Enfermería y
Fisioterapia - Conference
Venue

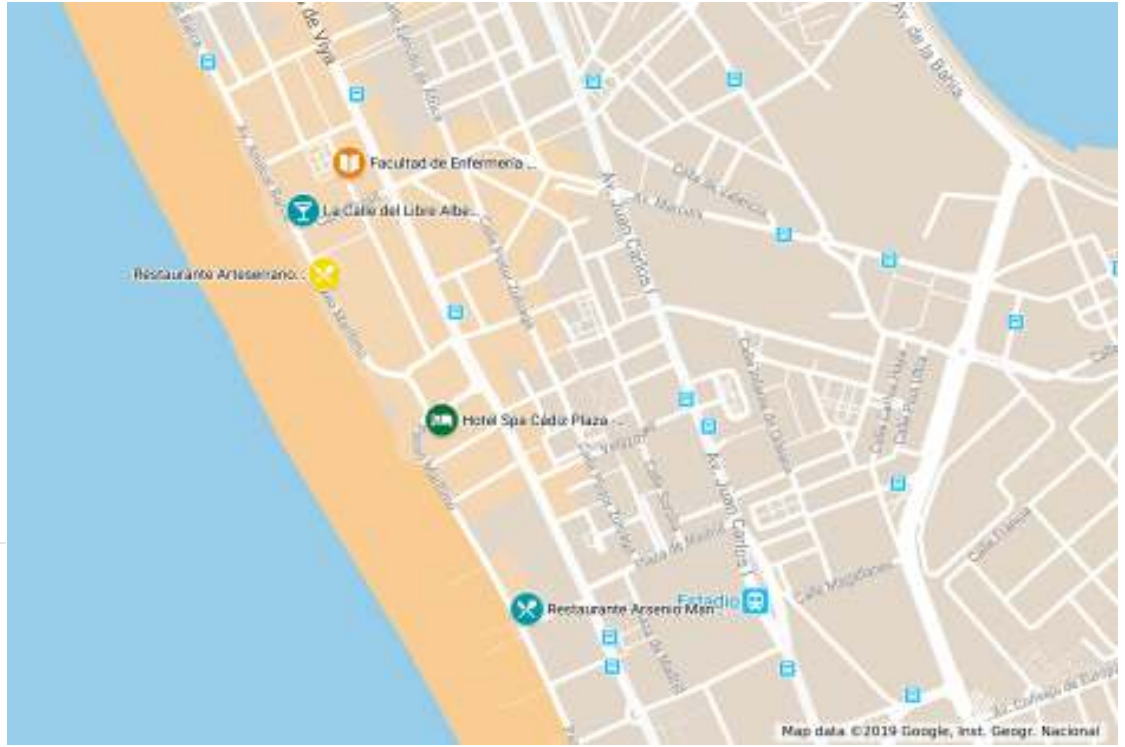

# PROCEEDINGS OF THE IX INTERNATIONAL WORKSHOP ON LOCATIONAL ANALYSIS AND RELATED PROBLEMS (2019)

Edited by

Marta Baldomero-Naranjo

Inmaculada Espejo-Miranda

Luisa I. Martínez-Merino
Juan Manuel Muñoz-Ocaña
Antonio M. Rodríguez-Chía

# Preface

The International Workshop on Locational Analysis and Related Problems will take place during January 30-February 1, 2019 in Cádiz (Spain). It is organized by the Spanish Location Network and Location Group GELOCA (SEIO). GELOCA is a working group on location belonging to the Statistics and Operations Research Spanish Society. The Spanish Location Network is a group of more than 140 researchers distributed into 16 nodes corresponding to several Spanish universities. The Network has been funded by the Spanish Government.

Every year, the Network organizes a meeting to promote the communication between its members and between them and other researchers, and to contribute to the development of the location field and related problems. Previous meetings took place in Segovia (September 27-29, 2017), Málaga (September 14-16, 2016), Barcelona (November 25-28, 2015), Sevilla (October 1-3, 2014), Torremolinos (Málaga, June 19-21, 2013), Granada (May 10-12, 2012), Las Palmas de Gran Canaria (February 2-5, 2011) and Sevilla (February 1-3, 2010).

The topics of interest are location analysis and related problems. It includes location models, networks, transportation, logistics, exact and heuristic solution methods, and computational geometry, among others.

The organizing committee.

## Scientific committee:

- María Albareda Sambola (U. Politécnica de Cataluña, España)
- Giuseppe Bruno (Università degli Studi di Napoli Federico II, Italia)
- Ángel Corberán (U. de Valencia, España)
- Elena Fernández Aréizaga (U. de Cádiz, España)
- Joerg Kalcsics (University of Edinburgh, UK)
- Martine Labbé (Université Libre de Bruxelles, Belgique)
- Alfredo Marín Pérez (U. de Murcia, España)
- Juan A. Mesa (U. de Sevilla, España)
- Stefan Nickel (Karlsrhue Institute of Technology, Germany)
- Justo Puerto Albandoz (U. de Sevilla, España)
- Antonio M. Rodríguez-Chía (U. de Cádiz, España)

## Organizing committee:

- Marta Baldomero Naranjo (U. de Cádiz)
- Inmaculada Espejo Miranda (U. de Cádiz)
- Luisa Isabel Martínez Merino (U. de Cádiz)
- Manuel Muñoz Márquez (U. de Cádiz, España)
- Juan Manuel Muñoz Ocaña (U. de Cádiz, España)
- Antonio M. Rodríguez-Chía (U. de Cádiz, España)
- Dolores Rosa Santos Peñate (U. de Las Palmas de Gran Canaria)
- Concepción Valero Franco (U. de Cádiz, España)

# Contents

PROGRAM

# Wednesday January 30th

## 16:00-16:30 Registration

## 16:30-16:45 Opening Session

## 16:45-18:25 Session 1: Discrete Location

Exact algorithm for the Reliability Fixed-Charge Location Problem with Capacity constraints
*M. Albareda-Sambola, M. Landete, J.F. Monge, and J.L. Sainz-Pardo*

An extension of the p-center problem considering stratified demand
*M. Albareda-Sambola, L.I. Martínez-Merino, and A.M. Rodríguez-Chía*

Capacitated Discrete Ordered Median Problems
*I. Espejo, J. Puerto, and A. M. Rodríguez-Chía*

Rationalizing capacities in the facility location problem
*Á. Corberán, M. Landete, J. Peiró, and F. Saldanha-da-Gama*

An exact algorithm for the Interval Transportation Problem
*M. Albareda-Sambola, M.Landete, and G. Laporte*

## 18:25-18:45 Coffee break

## 18:45-20:05 Session 2: Networks

Optimal allocation of fleet frequency for "skip-stop" strategies in transport networks.
*J. A. Mesa, F. A. Ortega, R. Piedra-de-la-Cuadra, and M. A. Pozo*

Dealing with Symmetry in a Multi-period Sales Districting Problem
*M. Bender, J. Kalcsics, A. Meyer, and M. Pouls*

Minmax Regret Maximal Covering on Networks with Edge Demands
*M. Baldomero-Naranjo, J. Kalcsics, and A. Rodríguez-Chía*

Non-dominated solutions for the bi-objective MST problem
*L. Amorosi and J. Puerto*

## 21:30 Welcome Reception

# Thursday January 31st

## 9:00-10:40 Session 3: Continuous Location

A Mixed Integer Linear Formulation for the Maximum Covering Location Problem with Ellipses
*V. Blanco and S. García*

Minimum covering polyellipses
*V. Blanco and J. Puerto*

The One-Round Voronoi Game Played on the Rectilinear Plane
*T. Byrne, S. P. Fekete, and J. Kalcsics*

Locating Hyperplanes for Multiclass Classification
*V. Blanco, A. Japón, and J. Puerto*

Introduction to planar location with orloca
*M. Muñoz-Márquez*

## 10:40-11:10 Coffee break

## 11:10-12:30 Invited Speaker: Francisco Saldanha da Gama

Logistics Network Design and Facility Location: The value of multi-period stochastic solutions

## 12:30-14:10 Session 4: Routing

The selective traveling salesman problem with time-dependent profits
*E. Barrena, D. Canca, L.C. Coelho, and G. Laporte*

Solidarity behavior for optimizing the waste selective collection
*E. Barrena, D. Canca, F. A. Ortega, and R. Piedra-de-la-Cuadra*

Steiner Traveling Salesman Problems: when not all vertices have demand
*J. Rodríguez-Pereira, E. Benavent, E. Fernández, G. Laporte, and A. Martínez-Sykora*

A Kernel Search for the Inventory Routing Problem
*C. Archetti, G. Guastaroba, D.L. Huerta-Muñoz, and M.G. Speranza*

A branch-and-price algorithm for the Vehicle Routing Problem with Stochastic Demands, Probabilistic Duration Constraint, and Optimal Restocking Policy
*A. M. Florio, R. F. Hartl, S. Minner, and J.J. Salazar-González*

## 14:10-15:30 Lunch

## 15:30-16:30 Session 5: Networks Design

Robust feasible rail timetable
*Á. Marín, M. A. Ruiz-Sánchez, and E. Codina*

Addressing locational complexity: network design and network rationalisation
*D. Ruiz-Hernandez, J. M. Pinar-Pérez, and M. B.C. Menezes*

Locating a new station/stop in a network based on trip coverage and times
*M. C. López-de-los-Mozos and J. A. Mesa*

## 16:40-17:40 Session 6: Networks Design II

The Urban Transit Network Design Problem
*A. De-los-Santos, D. Canca, A. G. Hernández-Díaz, and E. Barrena*

Infrastructure Rapid Transit Network Design Model solved by Benders Decomposition
*N. González-Blanco and J. A. Mesa*

The Railway Rapid Transit Network Construction Scheduling Problem
*D. Canca, A. de los Santos, G. Laporte, and J. A. Mesa*

## 17:40-18:00 Coffee break

## 18:00-19:00 Session 7: Applications

Wildfire Location Model: A new proposal
*J. A. Mesa and M. Marcos-Pérez*

Time dependent continuous optimisation in solar power tower plants
*T.Ashley, E. Carrizosa, and E. Fernández-Cara*

On computational Dynamic Programming for minimizing energy in an electric vehicle
*E. M.T. Hendrix* and I. Garcia

## 19:10-20:10 Session 8: Discrete Location II

Heuristic Framework to Reduce Aggregation Error on Large Classical Location Models
*C. Castañeda and D. Serra*

Emergency Vehicle Location Model considering uncertainty and the hierarchical structure of the resources
*J. Nelas and J. Dias*

Using a kernel search heuristic to solve a sequential competitive location problem in a discrete space
*D. R. Santos-Peñate, C. M. Campos-Rodríguez, and J. A. Moreno-Pérez*

## 21:30 Dinner

# Friday February 1st

## 9:00-10:40 Session 9: Applications/Routing/Hub Location

Minimum distance regulation and entry deterrence through location decisions
*J. Elizalde Blasco and I. Rodríguez Carreño*

The impact of pharmacy deregulation process on market competition and users' accessibility. Insights from two Spanish case studies.
*I. Barbarisi, G. Bruno, M. Cavola, A. Diglio, J. Elizalde Blasco, and C. Piccolo*

Solving the Ordered Median Tree of Hubs Location Problem
*M. A. Pozo, J. Puerto, and A. M. Rodríguez-Chía*

Feasible solutions for the Distance Constrained Close-Enough Arc Routing Problem
*M. Reula, Á. Corberán, I. Plana, and J. M. Sanchis*

Drone Arc Routing Problems
*J. F. Campbell, Á. Corberán, I. Plana, and J. M. Sanchis*

## 10:40-11:10 Coffee Break

## 11:10-12:30 Invited Speaker: Ivana Ljubic

Solving Very Large Scale Covering Location Problems using Branch-and Benders-Cuts

## 12:30-13:50 Session 10: Bilevel Location

On location-allocation problems for dimensional facilities
*L. Mallozzi, J. Puerto, and M. Rodríguez-Madrena*

New bilevel programming approaches to the location of controversial facilities
*M. Labbé, M. Leal, and J. Puerto*

A multi-period bilevel approach for stochastic equilibrium in network expansion planning under uncertainty
*L. F. Escudero, J. F. Monge, and A. M. Rodríguez-Chía*

Bilevel programming models for multi-product location problems
*S. Dávila, M. Labbé, F. Ordoñez, F. Semet, and V. Marianov*

**13:50-14:30 Location Network Meeting**

**14:30-15:30 Lunch**

INVITED SPEAKERS

# Logistics Network Design and Facility Location: The value of multi-period stochastic solutions

Francisco Saldanha da Gama,[1]

[1]*Universidade de Lisboa, Centro de Matemática, Aplicações Fundamentais e Investigação Operacional, Lisboa, Portugal*   faconceicao@fc.ul.pt

In the past decades logistics network design has been a very active research field. This is an area where facility location and logistics are strongly intertwined, which is explained by the fact that many researchers working in Logistics address very often location problems as part of the strategic/tactical logistics decisions. Despite all the work done, the economic globalization together with the emergence of new technologies and communication paradigms are posing new challenges when it comes to developing optimization models for supporting decision making in this area. Dealing with time and uncertainty has become unavoidable in many situations.

In this presentation, different modeling aspects related with the inclusion of time and uncertainty in facility location problems are discussed. The goal is to better understand problems that are at the core of more comprehensive ones in logistics network design. By considering time explicitly in the models it becomes possible to capture some features of practical relevance that cannot be appropriately captured in a static setting; by considering a stochastic modeling framework it is possible to build risk-aware models. Unfortunately, the resulting models are often too large and thus hard to tackle even when using specially tailored procedures. This raises a question: is there a clear gain when considering a more involved model instead of a simplified one (e.g. deterministic or static)? In search for an answer to this question, several measures are discussed that include the value of a multi-period solution and the value of a risk-aware solution.

# Solving Very Large Scale Covering Location Problems using Branch-and-Benders-Cuts

Ivana Ljubic[1]

[1] *ESSEC Business School of Paris, Cergy Pontoise Cedex, France*   ivana.ljubic@essec.edu

Covering problems constitute an important family of facility location problems with widespread applications. These problems embed a notion of proximity (or coverage radius) that specifies whether a given demand point can be served or "covered" by a potential facility location. Proximity is often defined in terms of distance or travel time between points. A demand point is then said to be covered by a facility if it lies within the coverage radius of this facility. Location problems with covering objectives or constraints are commonplace in the service sector (schools, hospitals, libraries, restaurants, retail outlets, bank branches) as well as in the location of emergency facilities or vehicles (fire stations, ambulances, oil spill equipments). They also find applications in the location of access points for the wireless communication in the smart grid deployments.

Many of these applications involve a relatively small number of potential facility locations while the number of demand points can run in the thousands or even millions. Such very large scale problem instances remain out of reach for modern MIP solvers.

In this talk we address the maximal covering location problem (MCLP) which requires choosing a subset of facilities that maximize the demand covered while respecting a budget constraint on the cost of the facilities and the partial set covering location problem (PSCLP) which minimizes the cost of the open facilities while forcing a certain amount of demand to be covered. We propose an effective decomposition approach based on the branch-and-Benders-cut reformulation. We also exploit the submodularity of the covering function and derive a formulation based on submodular cuts. A connection between Benders and submodular cuts is drawn as well.

The results of our computational study demonstrate that, thanks to decomposition, optimal solutions can be found very quickly, even for benchmark instances involving up to twenty million demand points.

The talk is based on a joint work with S. Coniglio, J.F. Cordeau and F. Furini.

ABSTRACTS

# An exact algorithm for the Interval Transportation Problem

Maria Albareda-Sambola,[1] Mercedes Landete,[2] and Gilbert Laporte[1,3]

[1]*Universitat Politècnica de Catalunya.BarcelonaTech, Terrassa, Spain,*  maria.albareda@upc.edu

[2]*Centro de Investigación Operativa,*
*Universidad Miguel Hernández de Elche, Spain,*  landete@umh.es

[3]*Canada Research Chair in Distribution Management, HEC Montréal, Montréal H3T 2A7, Canada*   Gilbert.Laporte@cirrelt.ca

This work focuses on the Interval Transportation Problem. For this problem, we propose two variants of an exact algorithm. Their efficiency is compared on a set of instances from the literature.

## 1.     Problem definition

Given a set of customers $j \in J$, each with demand $d_j > 0$, a set of facilities $i \in I$, each with capacity $q_i > 0$ and unit transportation costs $c_{ij}$ for $i \in I, j \in J$, the well known Transportation Problem (TP) consists in determining the amount of product to send from each facility to each customer so that each customer receives exactly his demand, the total amount delivered from each facility does not exceed its capacity, and the total transportation cost is minimized.

In the Interval Transportation Problem (ITP) it is assumed that, in fact, each demand can lie in a given interval: $d_j \in [\underline{D}_j, \bar{D}_j]$ and each capacity also: $q_i \in [\underline{Q}_i, \bar{Q}_i]$. Let $D = [\underline{D}_1, \bar{D}_1] \times \ldots \times [\underline{D}_m, \bar{D}_m]$, $Q = [\underline{Q}_1, \bar{Q}_1] \times \ldots \times [\underline{Q}_n, \bar{Q}_n]$, and $R = \{(d,q) \in D \times Q : \sum_{i \in I} q_i \geq \sum_{j \in J} d_j\}$ (Set of $(d,q)$ pairs with all demands and capacities within their intervals, that define a feasible TP instance) . Then, the goal of the ITP is to find

$$U^* = \max_{(d,q) \in R} z(d,q), \tag{1}$$

where $z(d, q)$ is the optimal solution of the TP defined by demands $d$ and capacities $q$. That is, the ITP aims at finding the feasible TP instance defined by parameters in $R$ with the most expensive optimal solution.

# 2.    Solution Algorithm

As already observed in [1], the ITP is in fact a bilevel optimization problem. For this reason, we propose to adapt to the ITP the algorithm presented in [2] for bilevel optimization problems among others. Roughly speaking, the authors propose a bisection line search on an interval containing the optimal value $U^*$; at each iteration a sequence of subproblems is solved to determine wether a given candidate value $u_0$ can be attained (there exists an instance defined by some $(d, q) \in R$ with $z(d, q) \geq u_0$) or not.

The work [2] considers a very general setting. For this reason, in that paper subproblems are solved by means of very general algorithms. In this work, we propose to take advantage of the structture of these subproblems in the particular case of the ITP to solve them more efficiently by suitably stating them as a linear or mixed integer linear problems.

Two alternative adaptations are proposed, that differ in the way how already visited TP solutions are considered. Their behavior is compared by means of a computational experience on a set of instances taken from the literature.

# Exact algorithm for the Reliability Fixed-Charge Location Problem with Capacity constraints

Maria Albareda-Sambola [1], Mercedes Landete [2], Juan Francisco Monge [2], Jose Luis Sainz-Pardo[2]

[1]*Departament de Estadística i Investigació Operativa,*
*Technical University of Catalonia-Barcelona, Terrassa, Spain,*
*maria.albareda@upc.edu*

[2]*Centro de Investigación Operativa,*
*Universidad Miguel Hernández de Elche, Spain,*
*landete@umh.es, monge@umh.es, jose.sainz-pardo@umh.es*

This work addresses the exact solution of the Reliability Fixed-Charge Location Problem with Capacity Constraints (RFLPCC). The proposed method is based on a formulation that considers all possible scenarios. Therefore, directly solving this model is computationally hard and it is not available to solve it even in small instances. We propose a dynamic approach in order to exactly bound the expected overload in the Reliability Fixed-Charge Location Problem. Too, we analyse by an exhaustive computational study the quality of the solutions according to several criteria.

## 1.    Dynamic approach

The dynamic approach proposed is based on the usual philosophy of master-slave problems. From the facilities that have been opened in the solution returned by the master problem, the slave problem solves a model in order to make a new assignment with expected overload under the given bound $B$. Both solutions are iteratively used to introduce different constraints in the master problem until the optimal value is obtained.

# 2. Computational experience

We compare by several computational experiments the dynamic approach versus the following three models proposed in [1] which can be seen as matheuristics for the RFLPCC:

- QRFLP: this model does not manage the overload, it only keeps the demand below the capacity of the facilities in the scenario in which no facility fails,

- CRFLP-B1: this model constrains an upper bound for the expected overload,

- CRFLP-LR: this model constrains a linear estimation of the expected overload, then it is an approximated model

Our intention is to analyse the performance of the proposed approach but not only in terms of optimal cost, also in terms of expected overload, overload probability, non-served demand, number of open facilities, computational time and instances solved before the limit time. Table 1 respectively contains the cited average values for one of the most representative computational set of instances of our study with a bound for the expected overload fixed to 3. Here, RFLP-EX stands for the exact algorithm presented in this work.

|  | $v^*$ | $E(X,Y)$ | P(overload) | Non-served | #Open | Time | Solved |
|---|---|---|---|---|---|---|---|
| QRFLP | 9806.13 | 5.4 | 0.12 | 0.22 | 3.6 | 349.25 | 20 |
| RFLP-B1(3) | 11090.57 | 1.32 | 0.10 | 2.44 | 4.4 | 1412.03 | 19 |
| RFLP-LR(3) | 10885.10 | 2.57 | 0.12 | 1.91 | 4.3 | 1337.10 | 17 |
| RFLP-EX (3) | 10496.21 | 2.84 | 0.15 | 1.17 | 4.1 | 1398.65 | 18 |

*Table 1.* Average values for a requested overload of $B$=3

# An extension of the $p$-center problem considering stratified demand [*]

Maria Albareda-Sambola,[1] Luisa I. Martínez-Merino,[2] and Antonio M. Rodríguez-Chía[3]

[1] *Universitat Politècnica de Catalunya.BarcelonaTech, Terrassa, Spain,* maria.albareda@upc.edu

[2] *Departamento de Estadística e Investigación Operativa, Universidad de Cádiz, Cádiz, Spain,* luisa.martinez@uca.es

[3] *Departamento de Estadística e Investigación Operativa, Universidad de Cádiz, Cádiz, Spain,* antonio.rodriguezchia@uca.es

This work introduces an extension of the classical discrete $p$-center problem ($pCP$), called the stratified $p$-center problem ($SpCP$). In this extension it is assumed that the population of each demand site is divided into different categories or strata depending on the kind of service that they require. The objective of the proposed model is to locate $p$ centers minimizing the weighted sum of the largest assignment distances associated with each stratum. This model could be applied in humanitarian relief planning where centers of humanitarian assistance cover different kind of needs and the demand of each need may be distributed in a spatially different way.

## 1.    Introduction to the problem

Given a set of potential locations $J = \{1, \ldots, n\}$ to open $p$ centers, a set of demand sites $I = \{1, \ldots, m\}$ and a set of strata $\mathcal{S}$, the goal of the $SpCP$

[*]Thanks to Agencia Estatal de Investigación (AEI) and the European Regional Development's funds (FEDER), projects MTM2013-46962-C2-2-P and MTM2016-74983-C2-2-R, Universidad de Cádiz, PhD grant UCA/REC02VIT/2014.

could be expressed as follows,

$$\min_{\substack{P \subseteq J \\ |P|=p}} \sum_{s \in \mathcal{S}} w_s d(I^s, P).$$

In this expression, $w_s$ is the weight related to stratum $s$, $I^s \subseteq I$ is the subset of demand sites where stratum $s$ is present and $d(I^s, P) = \max_{i \in I^s} \min_{j \in P} d_{ij}$. In addition, note that for a given site $i \in I$, we will refer to $\min_{j \in P} d_{ij}$ as the allocation distance of site $i$. Consequently, $d(I^s, P)$ is the maximum allocation distance among the sites with presence of stratum $s$.

In this work, we propose different formulations to address the $SpCP$. One of them is based on the approach proposed in [1] for the classical $pCP$. Besides, several formulations based on covering variables are introduced. They are inspired in formulations for classical discrete location problems as the ones appearing in [2] and [4].

In addition, we use the formulations and improvements developed for the $SpCP$ to obtain a heuristic approach for another extension of the $pCP$: the probabilistic $p$-center problem ($PpCP$), see [5]. This heuristic is based on the Sample Average Approximation (SAA), which is usually applied in discrete stochastic problems, see [3].

# Non-dominated solutions for the bi-objective MST problem[*]

Lavinia Amorosi[1] and Justo Puerto[2]

[1]*Dep. Statistical Sciences, Sapienza University of Rome, Italy,*  lavinia.amoros@uniroma1.it

[2]*IMUS, Universidad de Sevilla, Sevilla, Spain,*  puerto@us.es

This paper presents a new two phase algorithms for the computation of the entire set of non-dominated solutions of the bi-objective minimum spanning tree problem.

## 1.     The model

In this work we focus on a particularly appealing problem: the bi-objective minimum spanning tree (BMST) problem that has applications in different contexts [4]. For example, in the energy industry, for planning efficient distribution systems or in the telecommunication sector. The BMST problem in its basic form can be formulated as follows.

Let $G = (N, E)$ be an undirected graph with node set $N$ and edge set $E$. Let $c_1$ and $c_2$ be two different cost vectors on the edge set. The bi-objective minimum spanning tree problem is defined as:

$$\min Cx = \min(\sum_{e \in E} c_e^1 x_e, \sum_{e \in E} c_e^2 x_e) \tag{1}$$

$$s.t. \sum_{e \in E} x_e = n - 1 \tag{2}$$

$$\sum_{e \in E(S)} x_e \leq \mid S \mid -1 \ \ \forall S \subseteq N, \ \ S \neq \emptyset \tag{3}$$

$$x_e \in \{0, 1\} \quad \forall e \in E \tag{4}$$

[*]This research has been partially supported by Spanish Ministry of Economía and Competitividad/FEDER grants number MTM2016-74983-C02-01.

In the multi-objective context, and thus in the bi-objective case, the feasible set in the decision space (or decision set) $X = \{x \in R^n : Ax = b, x \geq 0\}$ is distinguished from the feasible set in the objective space (or outcome set) $Y = \{Cx : x \in X\}$, containing the points associated with the feasible solutions by means of the linear mapping defined by the problem criteria. Among the feasible solutions, we search for the ones which correspond to points in the outcome set for which it is not possible to improve one component without deteriorating another one.

The main contributions of this paper can be summarized as follows: 1) it provides a new approach to solve the bi-objective MST problem, based on a generic two-phase algorithm applicable to many bi-objective combinatorial optimization problems defined on graphs, 2) it gives a comparison between two alternative methods and open source solvers adopted for implementing the first phase: the dual variant of Benson's algorithm, [3], by means of BENSOLVE [5] and the weighted sum method by means of PolySCIP [2]; 3) it proposes a new enumerative recursive procedure based on the analysis of reduced costs, first introduced in [1] for the bi-objective integer min cost flow problem, able to generate all the spanning trees of a connected graph and it reports extensive computational results obtained testing the algorithm on different problem instances, including complete and grid graphs.

# A Kernel Search for the Inventory Routing Problem

C. Archetti[1], G. Guastaroba[1], D.L. Huerta-Muñoz[1], and M.G. Speranza[1]

[1]*Department of Economics and Management*
*Università degli Studi di Brescia. Brescia, Italy.*
claudia.archetti@unibs.it, gianfranco.guastaroba@unibs.it
diana.huertamunoz@unibs.it, grazia.speranza@unibs.it

In this talk, we propose a Kernel Search heuristic to solve the Inventory Routing Problem (IRP). The idea behind Kernel Search is to iteratively solve small restricted MILPs by selecting an initial subset of promising variables, called Kernel, and by adding sequentially subsets of the remaining variables, which are divided in groups of specific size, called Buckets. Preliminary results on a small set of benchmark instances show that the algorithm is able to improve, on average, the best-known solutions available in the literature.

## 1. The Kernel Search for the IRP

The *Inventory Routing Problem* [2] includes periodic demands, inventory management, and delivering–scheduling decisions over a given time horizon. The objective is to determine the best distribution plan over the time horizon to serve customers taking into account their consumption rate and inventory levels. The IRP variant we have focused this work takes into account a finite time horizon, a single depot that serves the customers, a fleet of homogeneous vehicles, a maximum-level policy, and the prohibition of stockouts.

The main contribution of this work is the development of an effective solution method, called *Kernel Search* (KS), to solve the IRP. KS is a general purpose scheme that has been proposed for the solution of Mixed-Integer Linear Programming (MILP) problems [4]. The idea is to identify a subset (or *Kernel*) of promising variables of the original problem and iteratively solve restricted MILPs by adding to this Kernel the remaining variables, which are grouped in small subsets called *Buckets*. To the best of our knowledge, KS has never been applied to IRPs.

Some preliminary experiments were run to analyze the performance of the KS in comparison with CPLEX and two state-of-the-art metaheuristics, M1 [1] and M2 [3], on a set of 72 benchmark IRP instances with a time horizon of six periods. In Table 1, we can observe that the KS obtained, on average, better results in considerably shorter computing times in most of the cases. The numbers inside the parentheses correspond to the number of solutions where KS outperforms the solution value found by the corresponding solution method.

*Table 1.* Preliminary results of the KS performance.

| Size | #Inst | Gap KS vs | | | Time(s) | | | |
|------|-------|-----------|-----------|-----------|---------|----------|----------|---------|
| | | CPLEX | M1 | M2 | CPLEX | M1 | M2 | KS |
| 10 | 12 | 0.19% | -0.87% | -12.63% | 7200 | 1871.17 | 11.02 | 3514.50 |
| 20 | 12 | 0.21% | -1.82% | -11.88% | 7200 | 2596.42 | 66.25 | 6885.50 |
| 30 | 12 | 0.48% | -0.06% | -11.09% | 7200 | 7224.75 | 222.57 | 6728.58 |
| 50 | 12 | -7.72% | 0.41% | -9.79% | 7200 | 11169.00 | 641.31 | 5448.17 |
| 100 | 12 | — | -1.05% | -10.66% | 7200 | 5474.75 | 2742.27 | 4831.92 |
| 200 | 12 | — | -0.81% | -10.68% | 7200 | 7801.17 | 16834.87 | 4852.17 |
| Average | | -1.72%(48) | -0.70%(49) | -11.13%(72) | 7200 | 6022.88 | 3419.72 | 5376.81 |

We expect these results can be significantly improved by carefully tuning the KS parameters (size of the initial kernel and buckets), which are a crucial issue as quality of the solution and computing times are strongly related to them.

# Time dependent continuous optimisation in solar power tower plants

Thomas Ashley,[1] Emilio Carrizosa,[2] and Enrique Fernández-Cara[2]

[1]*Instituto de Matemáticas de la Universidad de Sevilla, Spain,*  tashley@us.es

[2]*Instituto de Matemáticas de la Universidad de Sevilla, Spain,*  ecarrizosa@us.es

[3] *Dep. EDAN and IMUS, Universidad de Sevilla, Spain,*  cara@us.es

Research into renewable energy sources has continued to increase in recent years, and in particular the research and application of solar energy systems. Concentrated Solar Power (CSP) used by a Solar Power Tower (SPT) plant is one technology that continues to be a promising research topic for advancement.

The chosen aiming point for the heliostats on the receiver surface will have an effect on the production of energy, as well as an effect on the lifetime of the materials used in the receiver surface, due to thermal stresses. Therefore, the aiming strategy used by an SPT plant is of importance when seeking to achieve the optimal energy production, whilst minimising risk of damage to components.

The aiming strategy used in recent research into the optimisation of SPTs assumes that all heliostats in the field aim at the centre of the receiver, see [5]. This assumption allows for easier computation of the flux distribution across the receiver surface and reduces complexity of the adjustment of the heliostats. Some research has been conducted where more complex aiming strategies are considered for different receiver types [4].

In previous research [2] the authors considered the optimal aiming strategy for a SPT plant, assuming a fixed grid of aim points on the receiver surface, with run times low enough to allow for near real-time updates to the aiming strategy over time. This was extended in [3] to consider a continuous optimisation technique, whereby the aiming strategy for a particular time point was optimised without restricting the location or number of aiming points on the receiver.

In this work, the method from [3] is extended to consider the optimal aiming strategy across time, using a dynamic optimisation algorithm with an objective function of the form:

$$\text{Maximise} \quad J(p) = \int_0^T G(t, p(t)) dt \tag{1}$$

Subject to $p \in P_{ad}$ where $P_{ad}$ is the subset of a Hilbert space $P$ defined by time dependent inequality constraints of the form:

$$p(t) \in R \ \ a.e.$$
$$p \in Q, \ M(p) \le e. \tag{2}$$

Here, $R$ is a bounded and closed convex set in an Euclidean space, $Q \subset P$ is a second Hilbert space, $M : Q \mapsto E$ is a regular mapping with values in the Euclidean space $E$ and $e \in E$.

In this work, we discuss the existence of solutions and their optimality conditions and develop two possible algorithms to solve the problem. These algorithms are implemented in Python, and their functionality demonstrated with numerical examples for the SPT plant in Sanlucar la Mayor, Seville [1].

# Minmax Regret Maximal Covering on Networks with Edge Demands [*]

Marta Baldomero-Naranjo,[1] Jörg Kalcsics,[2] and Antonio M. Rodríguez-Chía[1]

[1]*Departamento de Estadística e Investigación Operativa, Universidad de Cádiz, Spain,* marta.baldomero@uca.es    antonio.rodriguezchia@uca.es

[2]*School of Mathematics, University of Edinburgh, United Kingdom,* joerg.kalcsics@ed.ac.uk

In this work, we focus our research on covering location problems. Almost all models analyzed in the literature assume that demand only occurs at the nodes of the network. However, there are some applications where this assumption is not realistic; e.g. the location of emergency facilities where the coverage areas are extremely distance-dependent. Thus, assuming that the demand is concentrated at nodes may lead to gaps in service levels that are not acceptable in some situations, rendering the solutions useless. Hence, our goal is to solve the single-facility location problem trying to cover the maximum demand on a network where the demand is distributed along the edges.

In the literature, see [1], some models are proposed to solve the deterministic version of this problem. Nevertheless, one of the big challenges is that the demand for a specific service is often not known exactly, but only approximately. Hence, we have to find locations for those facilities that provide an adequate level of service even under changing and unknown service demands. For this reason, we will treat demands as being unknown. However, we usually have a good idea of what the minimal or maximal demand will be, so that we can at least assume demand to lie within a known range. In the face of this situation of total uncertainty in the demand, we propose to employ concepts from robust optimization, more concretely minimizing the maximal regret, a well-known criterion used by many researchers, see e.g. [2].

---

[*]Thanks to the support of Agencia Estatal de Investigación (AEI) and the European Regional Development's funds (FEDER): project MTM2016-74983-C2-2-R, Universidad de Cádiz: PhD grant UCA/REC01VI/2017, Telefónica and the BritishSpanish Society Grant.

Our first aim is to provide mathematical models considering that demand is uncertain and distributed along the edges of a network and that the service facilities can, essentially, be located anywhere along the network. Furthermore, we will propose polynomial time algorithms for finding the location that minimizes the maximal regret assuming that the demand lies within a known range and it is constant or linear on each edge.

# The impact of pharmacy deregulation process on market competition and users' accessibility. Insights from two Spanish case studies.

Ilaria Barbarisi,[1] Giuseppe Bruno,[1] Manuel Cavola,[1] Antonio Diglio,[1] Javier Elizalde Blasco,[2] and Carmela Piccolo[1]

[1]*Department of Industrial Engineering (DII), Università di Napoli Federico II*
*P.le Tecchio, 80 - 80125, Napoli, Italy*
ilaria.barbarisi@libero.it, (giuseppe.bruno, manuel.cavola, antonio.diglio, carmela.piccolo)@unina.it

[2]*Facultad de Ciencias Económicas y Empresariales, Universidad de Navarra, Spain*
jelizalde@unav.es

Most European countries adopt regulations of the retail pharmacy market with the aim of guaranteeing some objectives in terms of accessibility, equity, efficiency and quality of services. In this regard, one critical aspect concerns the conditions for opening new pharmacies in a given area [1]. These conditions typically combine *demographic* (e.g. maximum number of pharmacies per inhabitant) and *geographic* (e.g. minimum distance among pharmacies) criteria, in order to ensure accessibility to medicine products for the entire population while preserving adequate market niches to the pharmacists. In the last years, many countries introduced policies aimed at promoting the competition in this sector. In particular, the restrictions for the release of licenses for new openings were progressively relaxed with a consequent increase in the number of opened pharmacies. In Spain, the Decree-Law 11/1996 established new threshold values valid at national level (i.e., minimum distance of 250 meters among pharmacies and one pharmacy per 2,800 inhabitants) but it transferred the right to the Autonomous Communities to modify such rules in order to better take into account the specificities of their competence areas. Some communities, like Catalunia, just implemented the indications fixed at national level while some others further relaxed the above values. The most relevant deregu-lation episode took place in 2000 in the region of Navarre, as the spatial and demographic criteria were respectively reduced to 150 meters and to one pharmacy per 700 inhabitants. These changes induced a dramatic entry process, almost doubling the overall number of pharmacies. [2].

In this context, the first aim of the present work is to evaluate the effects produced by this phenomenon, in terms of users' accessibility and cannibalization of potential customers among pharmacies. To this end, we selected two case studies, i.e. two cities belonging to different Autonomous Communities, and we performed an in-depth spatial analysis with the support of Geographic Information Systems (GIS) to represent demand points, facilities and to study their interaction.

Our analysis shows that the deregulation process produced effects that should be better addressed. On one side, the location of new pharmacies produced an overall increase of accessibility, but it has not contributed to make the access more equitable, as it has not improved the condition of the least well served users. On the other side, the cannibalization effect produced unbalanced situations, in which old pharmacies have not been able to maintain an adequate market niche. In this context, policy-makers are recommended to take actions to ensure equitable accessibility [3] and sustainable competition in a more deregulated environment [4]. To this end, more effective regulation mechanisms should be defined. We propose a mathematical model, aimed at generating alternative scenarios, with the objective of providing users with more equitable accessibility conditions to the service and, at same time, of mitigating the cannibalization effect among drugstores. The model is tested on the two selected real case studies. Obtained results show that the model is able to produce good scenarios, that can be evaluated by the Local Authorities to guide an informed-process for the definition of alternative regulation mechanisms.

# The selective traveling salesman problem with time-dependent profits[*]

Eva Barrena[1], David Canca[2], Leandro C. Coelho[3] and Gilbert Laporte[4]

[1] *University Pablo de Olavide, Seville, Spain*   ebarrena@upo.es

[2] *University of Seville, Seville, Spain*   dco@us.es

[3] *Université Laval, Québec, Canada,*   Leandro.Callegari-Coelho@fsa.ulaval.ca

[4] *Canada Research Chair in Distribution Management and HEC Montréal, Montréal, Canada,* gilbert.laporte@cirrelt.ca

Based on the definition of the selective traveling salesman problem (STSP), we define and analyze the selective travelling salesman problem with time-dependent profits (STSP-TDP). Given a weighted graph with time-dependent profits associated with the vertices, the STSP-TDP consists of selecting a simple circuit of maximal total profit, whose length does not exceed a pre-specified bound and whose starting and ending time must lie within a pre-specified planning horizon. The length of the planning horizon is bigger than the length of the circuit, thus being the starting and ending times of the circuit variables of the problem. This problem arises for example in the planning of tourist itineraries and in the collection of letters from mailboxes. We analyze several variants of the problem depending on the shape of the time-dependent profit functions. If these functions are not monotone, it may be worth visiting a site more than once. We propose a formulation for the case of multiple visits which reduces the problem to an STSP. We also propose three mathematical formulations for the single-visit case and compute optimal solutions for some benchmark instances.

[*]This research work was partially supported Ministerio de Economía y Competitividad (Spain)/FEDER under grant MTM2015-67706-P and by the Natural Sciences and Engineering Research Council of Canada (NSERC) under grant 2015-06189

# Solidarity behavior for optimizing the waste selective collection

Eva Barrena,[1] David Canca,[2] Francisco A. Ortega [3]
and Ramón Piedra-de-la-Cuadra[4]

[1]*Universidad Pablo de Olavide, Sevilla, Spain,*   ebarrena@upo.es

[2]*Universidad de Sevilla, Sevilla, Spain,*   dco@us.es

[3]*Universidad de Sevilla, Sevilla, Spain,*   riejos@us.es

[4]*Universidad de Sevilla, Sevilla, Spain,*   rpiedra@us.es

The problem of managing selective collection of waste within containers inside historic centers can be performed in three sequential phases: first, the location of containers along the streets; then, the determination of the minimum fleet size required to perform all collecting services; and finally, a model devoted to identify the optimal routes, in terms of total and equilibrated number of kilometers travelled by the trucks, is required. Obviously, the result of the first phase (location of the containers) highly influences the procedure since this will determine the decision to be taken for the subsequent phases (route of collection vehicles and service programming).

The main contribution of this paper focuses on this first phase: the location of collecting facilities (waste containers), where facility-customer distances must be considered in the collecting design system, as well as other considerations such as the size of container groups, their capacities in accordance with the closest population and the installation cost of those containers in specific sites along the streets.

On the other hand, we assume that customers are willing to have a solidarity behavior when they bring their trash bags. This behavior consists of using the container assigned to them within a pre-established proximity radius, although that container is not necessarily the closest to their place of residence. In this scenario, we show that a more efficient distribution of the containers can be obtained.

The proposed methodology for the deployment of containers for selective collection of urban solid waste can be identified as a version of the Partial Set Covering problem, whose computational complexity motivates the use of heuristics to face large real-life scenarios. Following that recommendation, a greedy algorithm of overflowing deviated to the immediate neighborhood has been developed to solve the proposed mathematical programming model.

To illustrate the performance of the developed methodology, a computational experience has been carried out on a network with randomized data inspired in a zone belonging to city of Seville (Spain).

# Dealing with Symmetry in a Multi-period Sales Districting Problem

Matthias Bender[1], Jörg Kalcsics[2], Anne Meyer[3] and Martin Pouls[1]

[1]*Department of Logistics and Supply Chain Optimization, Research Center for Information Technology (FZI), Haid-und-Neu-Str. 10–14, 76131 Karlsruhe, Germany*

[2]*School of Mathematics, University of Edinburgh, James Clerk Maxwell Building, The Kings Buildings, Edinburgh, EH9 3FD, Scotland, United Kingdom*

[3]*Faculty of Mechanical Engineering, TU Dortmund University, Leonhard-Euler-Straße 5, 44227 Dortmund, Germany*

In sales districting, the task is to assign a given set of customer accounts, each with a fixed market potential, to the individual members of a sales force such that each customer has a unique representative, each sales person faces an equitable workload and has an equal income opportunity, and travel times are minimal. Concerning the latter, if a sales person visits each customer every day, then the travel time is proportional to the length of a TSP tour. However, the workload of districts is usually balanced over several weeks and some customers may have to be visited only once during this time whereas others require weekly service. Moreover, customers may have time windows, tours may include overnight stays, and so on, which prohibits the computation of the actual travel times for practical problem sizes. Hence, in most cases one has to rely on estimates. The most common estimate is to compute either the sum of distances between a sales person's location and his assigned customers or the sum of pairwise distances between all customers assigned to the sales person.

One important, but only recently addressed aspect of sales districting is that customers often require service with different frequencies. Some customers have to be visited weekly, while others require service only once per month. As a result, planners not only have to design the districts, but also schedule visits to customers within the planning horizon. For example, if the planning horizon is divided into weeks and days, then we also have to decide which customers should be visited in which week and on which day of that week. This introduces a scheduling component to the district-ing problem. The criteria for scheduling customer visits are very similar to the ones for designing the sales territories. The total workload incurred by all customers served in each time period should be the same across all periods and the set of all customers visited in the same time period should be as compact as possible to reduce travel times during each period.

In this talk, we review the mixed-integer programming formulation for the problem that was derived in [1, 2]). Unfortunately, only very small instances can be solved to optimality within a reasonable amount of time with this formulation. One of the main factors contributing to that is, apparently, the high amount of symmetry that comes from the scheduling component. Therefore, we will present in this talk a characterisation of (some of) the symmetries arising in the problem and try to find some techniques to counter them. All findings will be supported by numerical tests.

# A Mixed Integer Linear Formulation for the Maximum Covering Location Problem with Ellipses

Víctor Blanco[1] and Sergio García[2]

[1]*Facultad de Ciencias Económicas y Empresariales, Universidad de Granada, Spain,* vblanco@ugr.es

[2]*School of Mathematics, University of Edinburgh, United Kingdom,* sergio.garcia-quiles@ed.ac.uk

In a covering location problem, there is a set of demand points and there is a set of potential sites for locating facilities. A point can be covered by a facility only if it is within a certain distance from this facility. Covering location problems have many applications in different areas like location of emergency services, analysis or markets, nature reserve selection, etc. In the Maximal Covering Location Problem (MCLP) introduced in [1], a fixed number of facilities must be located so that the amount of covered demand is maximized. As Euclidean distances on the plane are used in the MCLP, the geometric shape used to cover the demand point is a circle.

A much less studied variant of this problem is to use not circles but ellipses to cover the points, which has applications to wireless telecommunications networks as shown in [2]. There is a finite set of demand points and there is a finite catalogue of ellipses. The centers of these ellipses can be located anywhere on the plane. There are profits for serving the demand points and costs for locating the ellipses. The goal is to maximize the net profit by not using more than a certain number of ellipses given beforehand. As the formulation that introduced in [2] is nonlinear, the authors propose a simulated annealing heuristic that can solve very small instances. This method is outperformed by the exact and heuristic algorithms proposed in [3], although their formulation is still nonlinear. Our contribution in this paper is to use some geometric properties of this problem to give for the first time a mixed integer linear formulation that can solver much larger instances much more efficiently, as it will be shown with a computational study.

# Locating Hyperplanes for Multiclass Classification

Víctor Blanco,[1] Alberto Japón,[2] and Justo Puerto[2]

[1]*IEMath-Granada, Universidad de Granada, Granada, Spain*   vblanco@ugr.es

[2]*IMUS, Universidad de Sevilla, Sevilla, Spain*   ajapon1@us.es   puerto@us.es

In this work we present a novel approach to construct multiclass clasiffiers by means of arrangements of hyperplanes. We propose different mixed integer non linear programming formulations for the problem by using extensions of widely measures for misclassifying observations.

## General description

Given a training sample $\{(x_1, y_1), \ldots, (x_n, y_n)\} \subseteq \mathbb{R}^p \times \{1, \ldots, k\}$ the goal of supervised classification is to find a separation rule to assign labels $(y)$ to data $(x)$, in order to be applied out of sample. We assume that a given number of linear separators (hyperpanes in $\mathbb{R}^p$) have to be built to obtain a partition of the space into polyhedral cells. Each of the subdivisions obtained with such an arrangement of hyperplanes will be then assigned to a label in $\{1, \ldots, k\}$, see [5]. In Figure 1, where colors represent the different labels of $(y)$, we can see two examples with 5 hyperplanes partitioning $\mathbb{R}^2$ reaching a perfect classification. The formulations are based on the Support Vector Machines paradigm in which a maximum separation between classes is desired and in which different measures for the misclassifying errors are considered. Also, for the sake of solving larger instances, different strategies are proposed for the dimensionality reduction of the MINLP problems. We have run a series of experiments over some well known multiclass datasets from UCI machine learning repository [4]. In those we have tried four different versions of our model, using hinge-loss and ramp-loss measures for evaluating errors, and combining these with the $\ell_1$ and $\ell_2$ norms for measuring distances. We compare the results ob-tained with some of the most popular multiclass SVM techniques: One Vs One [1], Weston-Watkins [2], and Crammer-Singer [3].

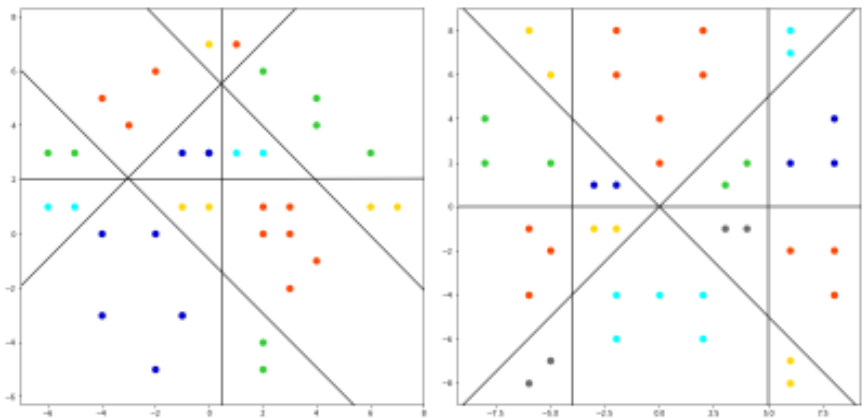

*Figure 1.*  5 hyperplanes multiclass SVM

# Minimum covering polyellipses

Víctor Blanco[1] and Justo Puerto[2]

[1]*IEMath-GR, Universidad de Granada, Spain,*  vblanco@ugr.es

[2]*IMUS, Universidad de Sevilla, Spain,*  puerto@us.es

In this work we study different continuous location problems using polyellipses. These problems allow to model situations in which one or several facilities are to be located in which the transportation costs are computed by means of the sums of distances to a set of sub-facilities.

## 1.     Introduction

Given a sets of demand points on the plane, Continuous Facility Location Problems (CFLP) deal with the determination of optimal positions by minimizing certain measures of the distances to the points. The most popular CFLP is the Weber problem [5], in which a finite set of demands points is provided, $\mathcal{U} \subseteq \mathbb{R}^2$, and a point $x \in \mathbb{R}^2$ is to be located that minimizes the function $\Phi(x) = \sum_{u \in \mathcal{U}} \omega_i \|x - u\|_2$, for some weights $\omega_1, \ldots, \omega_n$ (here $\| \cdot \|_2$ stands for the Euclidean norm). If the data points are not collinear, $\Phi$ is strictly convex, and therefore has a unique optimum, which can be obtained with the Weiszfeld's Algorithm [6]. The levels curves of $\Phi$ are given by the following sets:

$$\mathbb{E}_r(\mathcal{U}) = \left\{ x \in \mathbb{R}^2 : \sum_{u \in \mathcal{U}} \omega_j \|x - u\|_2 = r \right\}$$

for a radius $r \geq 0$. The set $\mathbb{E}_r(\mathcal{U})$ is called a polyellipse with foci $\mathcal{U}$ and radius $r$. The region bounded by $\mathbb{E}_r(\mathcal{U})$ is clearly a nonempty convex set provided that $r$ is greater than the optimal value of the Weber problem. These convex bodies have been partially analyzed from geometric or algrebraic viewpoints (see [1,3,4])

In this work, we analyze full covering problems by using these sets. On the one hand, we analyze covering problems, in which the goal is to find, for a given set of foci, $\mathcal{U}$ the smallest radius $r$ for which the polyellipse $\mathbb{E}_r(\mathcal{U})$ contains a set of given demand points, $\mathcal{A}$. A direct implication of this problem in facility location consists of the placement of the facilities in $\mathcal{U}$ (whose relative positions on the plane are given) such that the maximum sum of the distances from a demand point to the facilities is minimized. The problem extends the classical planar center problem [2,7] since polyellipses with a single foci are circles. Several approaches are provided to efficiently solve the problem: a Second Order Cone Programming formulation, a primal-dual approach, and a Elzinga-Hearn based algorithm. On the other hand, instead of providing the set of foci, $\mathcal{U}$, we propose a model and solution approaches to select, among the set of demand points, the relative positions of the foci to adequately cover the set of points. In Figure 1 we show the solutions of minimum radius polyellipses for different number of foci for the same dataset.

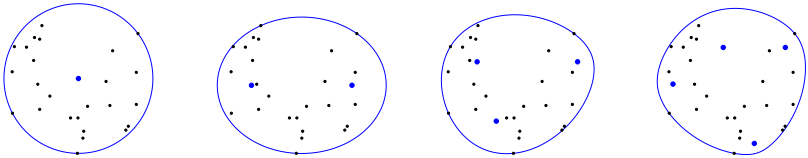

*Figure 1.* Full Covering Polyellipses for different number of foci.

# The One-Round Voronoi Game
# Played on the Rectilinear Plane

Thomas Byrne,[1] Sándor P. Fekete,[2] and Jörg Kalcsics[1]

[1] *University of Edinburgh, School of Mathematics, James Clerk Maxwell Building, King's Buildings, Edinburgh, EH9 3FD, United Kingdom*
tbyrne@ed.ac.uk  |  joerg.kalcsics@ed.ac.uk

[2] *TU Braunschweig, Mühlenpfordtstraße 23, 38106 Braunschweig, Germany*
s.fekete@tu-bs.de

Location is undoubtedly one of the most important issues when determining the success or failure of an operation. The distance between a proposed facility placement and its potential customer sites is perhaps the most natural way to discern the value of this position, and the need for effective facility locations becomes vital in competitive situations wherein customers will be gained or lost depending on whichever facility is closest. This importance of good location strategies is epitomised in the Voronoi game, a simple geometric model proposed in [1].

We consider this competitive facility location problem with two players. Players alternate placing points into the playing arena, until each of them has placed $n$ points. The arena is then subdivided according to the nearest-neighbour rule, and the player whose points control the larger area wins.

While some literature on this problem exists, there is a noticeable absence in the presentation of the game using the $l_1$ norm. A winning strategy for the second player, where the arena is a circle or a line segment, is presented in [1] for both variations where players can play more than one point at a time. There it is shown that the first player can ensure that the second player wins by an arbitrarily small margin. Optimal strategies for both players were found for a rectangular arena with Euclidean distance in [2] and it was ascertained that the particular values of $n$ and the aspect ratio of the arena determine which player wins.

We start with a definition of the game. There are two players, White and Black, each having $n$ points to play, where $n > 1$. The players alternate placing points on a rectangular playing area $\mathcal{P}$. As in chess, White starts the game, placing their first batch of points within $\mathcal{P}$, and Black the second batch of points, White the third batch, etc., until all $2n$ points are played. We assume that points cannot lie upon each other. Let $W$ be the set of white points and $B$ be the set of black ones. After all of the $2n$ points have been played, the arena is partitioned into the Voronoi diagram of $W \cap B$ using the $l_1$ metric and each player receives a score equal to the area of the Voronoi cells of their points, or rather their total market share.

The question we ask is *what is each player's best strategy?*

We answer this question in the one-round game, determining whether it is still chivalrous to play last, or is first the worst, second the best.

# Drone Arc Routing Problems

James F. Campbell,[1] Ángel Corberán,[2] Isaac Plana[3] and José M. Sanchis[4]

[1]*University of Missouri-St. Louis, USA,*  campbell@umsl.edu

[2]*Universitat de València, Spain,*  angel.corberan@uv.es

[3]*Universitat de València, Spain,*  isaac.plana@uv.es

[4]*Universidad Politécnica de Valencia, Spain,*  jmsanchis@mat.upv.es

In this talk we present some drone arc routing problems (Drone ARPs) and study their relation with well-known postman arc routing problems. Applications for Drone ARPs include traffic monitoring by flying over roadways, infrastructure inspection such as by flying along power transmission lines, pipelines or fences, and surveillance along linear features such as coastlines or territorial borders. Unlike the postmen in traditional arc routing problems, drones can travel directly between any two points in the plane without following the edges of the network. As a consequence, a drone route may service only part of an edge, with multiple routes being used to cover the entire edge. Thus the Drone ARPs are continuous optimization problems with an infinite number of feasible solutions. In order to solve them as a discrete optimization problem, we approximate each curve in the plane by a polygonal chain, thus allowing the vehicle to enter and leave each curve only at the points of the polygonal chain. If the capacity of the vehicles is unlimited, the resulting problem is a Rural Postman Problem (RPP). We propose an algorithm that iteratively solves RPP instances with an increasing number of points of the polygonal chain and present results on several sets of instances. We also briefly discuss the case in which the drones have limited capacity and several drones are needed.

# The Railway Rapid Transit Network Construction Scheduling Problem

David Canca,[1] Alicia de los Santos,[2] Gilbert Laporte[3] and Juan A. Mesa[4]

[1]*Department of Industrial Engineering and Management Science, University of Seville, Spain,*  dco@us.es

[2]*Department of Statistics, Econometrics, Operational Research, Management Science and Applied Economics, University of Cordoba, Spain,*  aliciasantos@uco.es

[3]*CIRRELT and HEC Montréal, Montréal, Canada,*  gilbert.laporte@cirrelt.ca

[4]*Department of Applied Mathematics II, University of Seville, Spain,*  jmesa@us.es

In this work we face up the problem of scheduling the different activities concerning the construction of a railway rapid transit transportation network. Supposing the network topology has been determined in an early stage, the problem consists on defining the best sequence of construction tasks in order to maximize the long term profit of the project. The problem can be viewed as a resource-constrained project scheduling problem, where both, construction budget and tunnel boring machines act as resources influencing the schedule. Since lines can be put into operation as they are finished, both, costs and revenues are dependent on the defined schedule.

## 1.     Introduction

The Resource-Constrained Project Scheduling Problem (RCPSP) is identified as the determination of the time required to implement the activities of a project to achieve a certain objective. It was assumed that the activities of a project are described by their processing times. Activities are related trough precedence relationships, which are commonly represented by sets of immediate predecessors. A certain amount of resources is required for each activity to be performed. It is assumed that, for each resource, a constant amount is available before the start of each period. The aim is to find the start time for all the activities accordingly to different objective functions: The minimization of the completion time [1], [5], the maximization of the net present value [2], [3] or the penalization of earliness-tardiness of the total completion time [4].

## 2. Contributions

The problem we face here presents some important differences with respect to the RCPSP. First, since transportation services can be put into operation as part of the lines are being finalized, the revenue is obtained as the project is being executed. A second difference is that there is not precedence constraints, instead, the temporal construction project is governed by connectivity constraints since, in normal circumstances, the transportation agency is interested in developing a connected network, allowing to passengers the possibility of transferring among lines. Finally, there is not a predefined initial task (or a set of initial tasks). The construction order will be a consequence of the city transportation demand patterns, which can vary along the time.

# Heuristic Framework to Reduce Aggregation Error on Large Classical Location Models

Carolina Castañeda P.[1] and Daniel Serra[1]

[1]*Department of Economics and Business, Universitat Pompeu Fabra, Barcelona, Spain,* carolina.castaneda@upf.edu, daniel.serra@upf.edu

Location analysis has a wide range of applications in the context of many real-world systems in public [2, 6] and private sectors [3], also it is a very interesting research field because its interaction with other disciplines [5, 7]. Facility location problems lie at the core of location analysis and are concerned with determining the location of a set of facilities satisfying one or more objective functions and constraints, regarding the demand for the service provided from the facilities.

Solving large discrete location problems may be time consuming or intractable due to the presence of many demand points, which are usually aggregated, thereby inducing error in the solution. According to [4], aggregation decreases costs of data modeling, collection and computing but also increases the errors incurred when solving location models to optimality because these are approximated models where the solution is optimal for the aggregated data but not necessarily for the non-aggregated data.

Aggregation error was first formally defined by [1]. Based on this definition the main classification of aggregation errors is the commonly known as ABC type errors [4]. Type A error appears when the distance between an aggregated demand point, instead of a real point, and a facility is used to solve a location problem. Type B error occurs when a facility is located at an aggregated demand point instead of a real point and type C error happens when a real demand point is assigned to the wrong facility.

We propose a heuristic framework to reduce error type C caused by aggregated demand points in classical location models on networks. Our framework integrates a solution method for large location problems with an algorithm to find a suitable demand aggregation for them.

The framework contains four stages. In the first stage, we obtain an initial aggregation of the original demand points through a heuristic based on the k-means algorithm. In the second stage, we calculate the centroid of each group, using the concept of a centroid in a minimum expansion tree. These centroids become the candidate locations to be selected by a location model solved in stage three. In the fourth stage, we evaluate the quality of the solution, calculating the improvement in the objective function with the current aggregation and then we repair the solution considering the aggregation error measure.

In this work we are avoiding type B error because facilities are located in the centroids that are real demand points. However, errors A and C are present. For measuring type C error, after finding the facility locations among centroids, we measure the dispersion of real demand points respect to the assigned facility, those points that have the largest dispersion are assigned to other facility that decrease the value of the total dispersion. Error A is not addressed in the current version of the framework.

The framework follows the four stages iteratively. After the first iteration, in the aggregation algorithm, we use a Greedy Randomized Adaptive Search Procedure (GRASP) in order to obtain new and diverse aggregated demand configurations. The algorithm stops when the solution quality does not improve marginally.

# Rationalizing capacities in the facility location problem

Ángel Corberán,[1] Mercedes Landete,[2] Juanjo Peiró[1] and Francisco Saldanha-da-Gama[3]

[1] Departament d'Estadística i Investigació Operativa. Universitat de València, Spain, angel.corberan@uv.es,juanjo.peiro@uv.es

[2] Departamento de Estadística, Matemáticas e Informática. Instituto Centro de Investigación Operativa. Universidad Miguel Hernández de Elche, Spain landete@umh.es

[3] Departamento de Estatística e Investigação Operacional. Centro de Matemática, Aplicações Fundamentais e Investigação Operacional. Universidade de Lisboa, Portugal, fsgama@fc.ul.pt

The capacitated facility location problem is a core problem in Location Science (see [3] for a survey of fixed-charge location problems). In this work we consider a variant of this problem in which facilities may cooperate in order to adapt their capacities to the demand of their customers. In particular, we consider the situation in which there may be capacity transfer between facilities. The existence of a potential flow between facilities leads to a redefinition of the capacity of a facility, and the actual capacity of a facility results from the original one plus the amount received from other facilities minus the amount sent to other facilities. This problem has multiple applications in real markets, see [1] and [2]. In this work we introduce the problem and give a mixed-integer linear formulation. Then, we enhance the model by using several families of valid inequalities. Some of the valid inequalities imitate classical valid inequalities of the capacitated facility location problem while some other are specific for this problem. Computational results illustrate the benefits of both models and their improvements.

# Bilevel programming models for multi-product location problems[*]

Sebastián Dávila,[1] Martine Labbé,[2] Fernando Ordoñez, [3]
Frédéric Semet,[4] and Vladimir Marianov,[5]

[1]*Department of Industrial Engineering , Universidad de Chile, Chile,*
sebastian.davilaga@gmail.com

[2]*Computer Science Department, Université Libre de Bruxelles, Belgium and INRIA, Lille, France,*
mlabbe@ulb.ac.be

[3]*Department of Industrial Engineering , Universidad de Chile, Chile,*
fordon@dii.uchile.cl

[4]*CRIStAL Centre de Recherche en Informatique Signal et Automatique de Lille, France, and INRIA, Lille, France,*
frederic.semet@centralelille.fr

[3]*Department of Electrical Engineering, Pontificia Universidad Católica de Chile, Santiago, Chile,*
marianov@ing.puc.cl

We consider a retail firm that owns several malls with a known location. A particular product, e.g., food processor, comes in $p$ types, which differ by shapes, brands and features. The set of all $p$ products is $P$. Each mall $j$ has a limited capacity $c_j$ of products in $P$ to be sold at that location, so the firm has to choose what products to sold at what mall. Furthermore, the firm can apply discrete levels of discount on the products, e.g., 5% and 10% over the price $\pi_k$ of product $k$. The objective of the firm is to find what products to sell at which mall, with what level of discount, so that its profit is maximized.

Consumers are located in points of the region. Each consumer or group of consumers $i$ has a different set $P_i \subseteq P$ of acceptable products, and will

[*]Supported by INRIA Associated Team BIPLOS

purchase one of these, or none if it is not convenient for her. Consumers maximize their utility, defined as

$$u_{ijkl} = r_{ik} - \alpha_{jkl} \cdot \pi_{jk} - 2d_{ij} \tag{1}$$

where $r_{ik}$ is the maximum expenditure that customer $i$ is willing to make to acquire product $k$; $\alpha_{jkl}$ is (100 - discount level $l$ in percent) of the product $k$ in mall $j$; $\pi_{jk}$ is the price of the product $k$ in the mall $j$ and $d_{ij}$ is the distance between consumer $i$'s origin and mall $j$. Whenever this utility is negative for product $k$ at all malls, the consumer does not purchase the product.

The agents (firm and consumers) play a Stackelberg game, in which the firm is the leader and the customers the follower. Once the firm decides the products to sell at each mall and the possible discounts, consumers purchase (or not) one of their acceptable products wherever their utility is maximized. We model the problem using first a bilevel formulation, and we further replace the follower problem by the primal constraints and optimality restrictions, to obtain a compact formulation. We also present a strong and a weak formulation.

# The Urban Transit Network Design Problem

Alicia De-Los-Santos,[1] David Canca,[2] Alfredo G. Hernández-Díaz[3] and Eva Barrena[4]

[1]*Department of Statistics, Econometrics, Operational Research, Management Science and Applied Economics, University of Cordoba, Spain,* aliciasantos@uco.es

[2]*Department of Industrial Engineering and Management Science, University of Seville, Spain,* dco@us.es

[3]*Deparment of Economic, Quantitative Methods and Economic History, University of Pablo de Olavide, Spain,* agarher@upo.es

[4]*Deparment of Economic, Quantitative Methods and Economic History, University of Pablo de Olavide, Spain,* ebarrena@upo.es

In this work we consider the problem of simultaneously designing the infrastructure of a urban bus transportation network and its set of lines while minimizing the total travel time of all passenger willing to travel in the network. As main differences with respect to other works in the bus transportation design field, we do not consider an a priori line pool, but we design the set of lines from square one, presenting a detailed description of the travel time (which incorporates the time spent in transferring along the passengers paths) and we jointly determine the transit assignment accordingly to the users' minimum trip time.

The transfer time plays an important role in the passenger decisions since transfers represent discomfort for the passenger, i.e., an extra-time to perform a trip. Most authors incorporate transfers into the computation of the travel time as a penalty term considering only the number of transfers. In this work we are interested in introducing a detailed description of the transfer time. To this end, we consider two layers: the first one affecting the off-board passengers movement (pedestrian layer) and the second one corresponding to the road infrastructure over which buses can run along (road-infrastructure layer).

In a realistic way, we can distinguish two types of transfers: transfers at the same stop and transfer between different stops. Obviously, the second type requires an extra-time to walk between stops over the pedestrian layer and therefore, a greater discomfort for passengers. We present a mathematical programming model for solving the problem on the directed graph that results when superimposing both layers.

We illustrate the problem with some computational experiments over several network.

# Minimum distance regulation and entry deterrence through location decisions

Javier Elizalde,[1] and Ignacio Rodríguez Carreño,[2]

[1]*University of Navarra, Pamplona, Spain,*   jelizalde@unav.es

[2]*University of Navarra, Pamplona, Spain,*   irodriguezc@unav.es

This paper analyses the location strategies and the resulting market structure in a model of spatial competition, illustrating location in two dimensions, when there is a restriction of minimum distance between plants. Such a regulation exists in some retail markets, such a drugstores, aiming to avoid agglomeration and provide accessibility for all consumers, and may change the optimal location decisions of managers with the purpose of reducing the eventual number of competitors. The latter becomes endogenously determined by the size of the market and the distance rule and we evaluate the welfare consequences of the firms' location strategies when they take their decisions with the purpose of deterring additional entry. We describe a theoretical model of spatial competition in two dimensions and solve for the equilibria through algorithmic simulations. The eventual number of active firms becomes endogenously determined by the size of the market and the distance rule. In a sequential entry game, we obtain a location equilibrium for a wide range of the binding distance rule and compare the equilibria reached under two types of firm behavior: with a simple maximum capture behavior and with entry deterrence strategies. We then discuss the results in terms of welfare in order to assess the effect of such regulatory policies and the distortions in firm's location decisions that they imply. The main finding of the paper is that, with a minimum distance constraint, location equilibrium exists for each level of minimum distance which allows for two or more firms. Even though entry deterrence activities by incumbent firms tend reduce the level of consumers' welfare as it tends to reduce the number of firms for some levels of minimum distance, it may in some cases be welfare enhancing as it may lead to a more even distribution of firms in the plane reducing the distance travelled by the average consumer.

# A multi-period bilevel approach for stochastic equilibrium in network expansion planning under uncertainty

Laureano F. Escudero,[1] Juan Francisco Monge,[2] and Antonio M. Rodríguez-Chía[3]

[1] *Area de Estadística e Investigación Operativa, Universidad Rey Juan Carlos, Spain,* laureano.escudero@urjc.es

[2] *Centro de Investigación Operativa, Universidad Miguel Hernández, Spain,* monge@umh.es

[3] *Departamento de Estadística e Investigación Operativa, Universidad de Cádiz, Spain,* antonio.rodriguezchia@uca.es

This study focuses on the development of a mixed 0-1 bilinear modeling stochastic framework for the multi-period network expansion planning problem under uncertainty, where stochastic equilibrium-based strategic decisions are to be made. The problem addressed here involves strategic decisions to be taken for a stochastic equilibrium-based network expansion planning problem (SE-NEP) in a multi-period time horizon under uncertainty, where it is required that the optimization models for the upper and lower levels have an equilibrium at the nodes of the scenario tree.

Some important features distinguish this work from other ones in the literature, they are as follows:

Topological decisions are now considered as dynamic decisions taken at different periods along a time horizon.

Several sources of uncertainty are considered at the periods, namely, investment cost for building network links, commodities volume to be transported by using network links as well as alternative ones, and commodities transport unit cost through alternative links, among others. The uncertainty is assumed to be captured by a finite set of scenarios.

A multi-period SE-NEP is dealt with, where some extensions of the classical static deterministic TAP (Toll Assignment Problem) are taken as a pi-lot case. On one hand, an upper level multi-period stochastic model for expansion planning is considered for determining the network links investment as well as the unit tariff for commodities transportation in order to maximize the expected profit in the scenarios along the time horizon. On the other hand, a lower-level single-period deterministic model is considered at each scenario node for minimizing the commodities transport cost in a mix of available network links and alternative ones. So, an equilibrium is sought for obtained on upper level profit and lower level cost at each scenario node. That equilibrium is obtained via the optimization of a single model. That type of modeling has been preferred to the also classical KKT constraint system due to computational reasons baswed on the model's difficulty for problem solving.

The upper level investment-oriented 0-1 step variables modeling objects allow that the state variables in the model only link two consecutive periods. A new feature of the problem that is considered consists of allowing upper bounded freedom for considering tariffs in the commodities transportation through the already available network links. As a consequence a related mixed 0-1 bilinear term is considered for each commodity transportation through the network links at each scenario node. Those terms are equivalently replaced with mixed 0-1 linear constraint systems.

Given the huge problem's dimensions (due to the network size of realistic instances as well as the cardinality of the scenario tree), it is unrealistic to seek for an optimal solution. As an illustrative example, for an instance with 20 network nodes, 12 network links, 25 links by other means, 4 periods, 156 scenario nodes and 125 scenarios to represent the uncertainty, the mathematically equivalent mixed 0-1 deterministic model has the following dimensions 38,1108 constraints, 84,396 continuous variables and 156,624 0-1 variables. This fact motivates the development of a matheuristic algorithm based on a Nested Stochastic Decomposition methodology for determining the appropriate network expansion planning along the time horizon, where a solution goodness quality is guaranteed. Some comutational results are shown.

This research has been partially supported by the projects MTM2015-63710-P (L.F. Escudero), MTM2016-79765-P (Juan F. Monge) and MTM2016-74983-C2-2-R (Antonio Rodríguez-Chía) from the Spanish Ministry of Economy, Industry and Competitiveness and the European Regional Development Fund (AEI/FEDER, UE).

# Capacitated Discrete Ordered Median Problems

Inmaculada Espejo,[1] Justo Puerto,[2] and Antonio M. Rodríguez-Chía[1]

[1]*Departamento de Estadística e Investigación Operativa, Universidad de Cádiz, Spain*

[2]*Departamento de Estadística e Investigación Operativa, Universidad de Sevilla, Spain*

Flexible discrete location problems, or the so-called discrete ordered median problems, have been widely studied in combinatorial optimization. In this paper we deal with the capacitated version of this problem. Different formulations of the capacitated discrete ordered median problem are presented as well as some preprocessing phases for fixing variables. In addition, some strategies to generate incomplete formulations and increase the size of the instances that we are able to solve are also studied.

## 1.    Introduction

Discrete location problems have been intensively studied over the last decades. Numerous surveys and textbooks give evidence of this fact (see [1, 2]). The need for location models that better fit different real world situations has made necessary to develop new and flexible location models. Nickel in [5] proposed the Discrete Ordered Median Problem (DOMP) which is used to model different discrete locations problems. It is a flexible formulation that introduce a type of objective function called ordered median function. This objective function is based on applying an ordered weighted averaging operator to the costs as they appear in the solution and taking them into account with a suitable vector $\lambda$. Hence, handling the most important objective functions in Location Analysis is possible with one unique model and also new ones may be created by adapting the parameters $\lambda$ adequately. Important references in location problems are [6,9], which summarize and review modeling and solution approaches published for continuous, network and discrete location problems.

This paper deal with the capacitated discrete ordered median problem (CDOMP). Flexible models using capacity constraints can be found in [3, 4, 7, 8]. In [8] provided formulations for the case of hubs. To the best of our knowledge, the first paper dealing with the capacitated discrete ordered median problem is [7]. It considers a new formulation based on a coverage approach and compare its performance with respect to previously known formulations for the uncapacitated problem. However, it is able to solve only small instances. In [4] describe three different points of view of a location problem in a logistics system. These models are extensions of the basic DOMP but the demands can be split. [3] can be considered as a first building block in the analysis of capacitated strategic location problems with order requirements. Here the number of facilities to be located is not given in advance. This is an important difference to the CDOMP, where the number of new facilities is fixed a priori.

# A branch-and-price algorithm for the Vehicle Routing Problem with Stochastic Demands, Probabilistic Duration Constraint, and Optimal Restocking Policy[*]

Alexandre M. Florio[1], Richard F. Hartl[1], Stefan Minner[2], and Juan-José Salazar-González[3]

[1] *University of Vienna, 1090 Vienna, Austria*

[2] *Technical University of Munich, 80333 Munich, Germany*

[3] *Universidad de La Laguna, 38271 Tenerife, Spain*   jjsalaza@ull.es

When customers' demands are stochastic, the duration of the routes are also stochastic. We consider the vehicle routing problem with stochastic demands and probabilistic duration constraints where we must design a set of routes with minimal total expected cost, visiting all customers, and such that the duration of each route, with some high probability, does not exceed some prescribed limit. We assume that the "Optimal Restocking Policy" is applied, which means that, before starting a route, the drive is given with a sequence of customers to serve, and with threshold values to check whether the vehicle must continue directly to the next customer in the sequence, or restock at the depot before. This is a more convenient and sophisticated policy than the "detour-to-depot policy", commonly used in the literature and where the driver is only given with the customer sequence and must keep following the route until it failures or ends at the depot. The problem, without duration constraint, has been recently studied in [1–3].

We solve the problem to optimality for the first time with a novel branch-and-price algorithm. An orienteering-based completion bound is proposed to control the growth of labels in the pricing algorithm. New procedures are developed to keep track of the variance of the duration of a route. The feasibility of an a-priori route is verified either by applying Chebyshev's bounds, or by Monte Carlo simulation and statistical inference. Consistency checks are incorporated into the branch-and-price framework to detect statistical errors. Computational experiments are performed with demands following binomial, Poisson, or negative binomial probability distributions, and with duration constraints enforced at levels of 90%, 95% and 98%. The vehicle capacity is considered in the objective function to force the vehicle going to the depot when convenient. Then, optimal solutions may contain a-priori routes that serve an expected demand larger than the capacity of the vehicle. These solutions actively employ optimal restocking to reduce traveling costs and the number of required vehicles when compared to the detour-to-depot policy solutions. Sensitivity analyses indicate that over-dispersed demands and strict duration constraints negatively impact the solution, both in terms of total expected cost and number of routes employed.

---

[*]Research partially supported by MTM2015-63680-R (MINECO/FEDER, UE)

# Infrastructure Rapid Transit Network Design Model solved by Benders Decomposition

Natividad González-Blanco[1] and Juan A. Mesa[2]

[1]*University of Seville, Seville, Spain,*   ngonzalez2@us.es

[2]*University of Seville, Seville, Spain,*   jmesa@us.es

In recent years, citizens mobility patterns are increasing due to longer trips which are caused by some factors as house spreading, enlargement of the urbanised areas, traffic problems in the city centres or in entrances of cities and the reduction of average ground traffic speed. These are some reasons why new rail transit systems have been constructed or expanded in determined agglomerations, or are being planned for construction. This investment is motivated by the necessity of energy saving and for pollution reduction too. Besides, in cases in which exist an infrastructure this is motivated by the increase in travel demand. The Rapid Transit Network Design Problem consists of locating alignments and stations covering as much as possible, knowing that the demand makes its own decisions about the transportation mode.

The infrastructure design problem has been treated in some papers as in [1].We propose some modifications of the model in order to improve the computational time for medium size networks. It is known that the model consists of maximizing trip coverage taking into account some considerations as budget limits and routing demand conservation. It is assumed that the mobility patterns in a metropolitan area are known and also, the locations of potential stations and potential links between each pair of stations. In addition, there already exists a different mode of transportation (for example, a bus network or private cars) competing with the railway to be built. According to other papers, we allocates the demand by using travel time.

As in previous models, this has a budget constraint, an alignment location constraint and a set of routing demand conservation constraints. In addition, it has an location-allocation constraint and an splitting demand constraint.

Actually this problem is difficult to solve because it has a lot of binary variables and, of course, constraints. Branch&Bound does not get to solve it in a efficient way. This reason has motivated us to applied Benders Decomposition algorithm and modifications of it. Subsequently, Branch&Bound and Benders Decompositions has been compared computationally.

# On computational Dynamic Programming for minimizing energy in an electric vehicle[*]

Eligius M.T. Hendrix[1], Inmaculada Garcia[1]

[1]*Computer Architecture, Universidad de Málaga*   eligius@uma.es    igarciaf@uma.es

In literature, one can find a branch and bound approach for the control of electric vehicles was published. Using that model, we create a DP implementation to obtain similar results.

## 1.     Modelling energy consumption

Literature on control typicially focuses on continuous control using theory about Hamiltonians and co-states. In contrast, [1, 2] applied a completely different approach based on branch and bound (B&B). Given our experience applying dynamic programming (DP), our hypothesis is that computational dynamic programming can reach the same result. This provides potential for further investigating the generation of control tables. Moreover, for the dynamics of the model, we will apply difference equations rather than differential equations based on a step size of 0.1 seconds and a time index $t = 0, \ldots, T$. For parameter values we use captial letters and for variable lower case symbols. Parameter values:

$F$: Final control horizon in seconds; $T = \frac{F}{\delta}$ : Number of periods in the horizon; $P$: Target position to be reached in control horizon; $R$: Radius of the wheels in m, $B = .05$ Ohm: Resistance of the battery; $S = 150$ volts: Voltage of power supply, $Tr = 10$: Transmission coefficient motor to wheels; $C = .517$: resistance depending on air density, surface car and aerodynamics; $L = .05$: Inductance rotor; $I = .03$ Ohm: Inductor resistance;

---
[*] This paper has been supported by The Spanish Ministry (TIN2015-66680-C2-2-R) in part financed by the European Regional Development Fund (ERDF).

$Q = .27$: Coefficient motor torque; $M = 250$ kg: Mass vehicle; $G = 9.81$: Gravity constant; $Fr = .03$: Friction of the wheels; $J$: Summarizing constant $J = \frac{Fr^2}{MR^2}$

Variables
$i_t \in [-150, 150]$    Induction of the engine
$\omega_t$    radial speed (radius/second), i.e. velocity $v_t = \frac{3.6R}{Tr}\omega_t$
$p_t \in [0, P]$    position of the vehicle
$u_t \in \{-1, 1\}$    Control, switch.

As, one can switch very frequently, $u_t$can also be considered continuous. We will make use of that to limit its value such that $i_t \in [-150, 150]$. The objective is given by

$$E = \sum_{t=0}^{T-1} Su_t i_t + Bu_t^2 i_t^2. \tag{1}$$

The dynamics is given by difference equations taking the time step size $\delta$ into account. Position:

$$p_t = p_{t-1} + \delta v_t. \tag{2}$$

Induction:

$$i_t = i_{t-1} + \delta \frac{Su_t - Ii_{t-1} - Q\omega_{t-1}}{L}. \tag{3}$$

To keep the induction into boundaries, we limit the control to

$$u_t \in \{\max\{\Delta_t, -1\}, \min\{\Delta_t, 1\}\}$$

with

$$\Delta_t = \frac{150L + (\delta I - L)i_{t-1} + \delta Q\omega_{t-1}}{\delta S}$$

Radial speed

$$\omega_t = \omega_{t-1} + \delta J(Qi_{m-1} - \frac{R}{Tr}(MGFr + Cv_{t-1}^2)). \tag{4}$$

The idea is to find the trajectory $u_t$ of control in order to minimise the total energy consumption. We will show how this can be realised applying DP.

# New bilevel programming approaches to the location of controversial facilities [*]

Martine Labbé,[1] Marina Leal,[2] and Justo Puerto[2]

[1]*Computer Sciences Department of the Université Libre de Bruxelles.*  mlabbe@ulb.ac.be

[2]*IMUS, Universidad de Sevilla.*  mleal3@us.es  puerto@us.es

We propose a novel bilevel model for the location of facilities whose placement generates disagreement among users with different, non-aligned or opposite interests. We develop the bilevel location model for one follower and for any polyhedral distance, and we extend it for several followers and any $\ell_p$-norm, $p \in \mathbb{Q}$, $p \geq 1$.

## 1. The models

Motivated by recent realworld applications in Location Theory in which the location decisions generate controversy, we propose a novel bilevel location model in which, on the one hand, there is a leader who wants to locate some primary facilities and must choose among a fixed number of potential locations where to establish them. Next, on the second hand, there is one follower that, once the primary facilities have been set, chooses the placement of a secondary facility, in a continuous framework. The leader and the follower have opposite targets; the leader's and follower's goal is to maximize and minimize, respectively, some proxy of the overall weighted distance between the primary and secondary facilities. We assume that the distances are measured via polyhedral distances. We prove then the NP-hardness of the models.

Later, we extend the model to the cases in which several followers are involved in the decision process and/or the considered distances are not polyhedral but more general distances induced by $\ell_p$-norms with $p \in \mathbb{Q}$, $p \geq 1$.

---

[*]This work has been partially supported by MINECO Spanish/FEDER grants number MTM2016-74983-C02-01.

Examples of this controversial location can be found in the literature, for example, in areas of semiobnoxious facilities, [1,3,4], or in the location and protection of critical infrastructures or facilities sensitive to intentional attacks, [2,5].

# 2. Mathematical programming formulations and resolution algorithms

In order to deal with the model with one follower and polyhedral distances we develop two different procedures: one based on the evaluation of the norm through its primal expression, and other based on the evaluation of the norm through its dual expression. For each of the procedures we develop Mixed Integer Linear Programming formulations, using duality, and also alternative Benders decomposition algorithms.

We conduct a computational study that shows the very-good performance of the Benders algorithms, being able to solve instances with $10000$ possibilities for the primary facilities from dimension $2$ until dimension $20$.

For the extension to several followers and $\ell_p$-norms we use replicas of the follower problem and conic duality, respectively.

# Locating a new station/stop in a network based on trip coverage and times

María Cruz López-de-los-Mozos,[1] and Juan A. Mesa,[2]

[1] *Dpto. Matemática Aplicada I. Universidad de Sevilla, Spain,*  mclopez@us.es

[2] *Dpto. Matemática Aplicada II. Universidad de Sevilla, Spain,*  jmesa@us.es

There are in the literature several covering location problems in a planar-network context (see a review in [2], and references therein). Some of them are devoted to cover origin-destination pairs (OD-pairs) instead of single points [1, 3]. This work is also focused on covering OD-pairs in a mixed transportation mode context, in which traveling times are a combination of planar and network times.

More specifically, we consider a network embedded in the plane representing a rapid transportation system, such that the nodes are either junctions or stations/stops already located, and we assume a set of existing facilities in the plane (not necessarily belonging to the network), such that traveling along the network is faster than traveling within the plane with some planar metric (in this work, the Euclidean metric). An OD-pair is said to be covered if the time spent in the combined plane-network mode is lower than in the planar mode. Within this context, the problem of locating a new station on the network is studied. The aim is to maximize the trip coverage when heterogeneous dwell times at the stations of the network are considered.

In order to incorporate considerations on trip times to the problem, some insights must be taken into account. First, locating a new station increases the accessibility of the network, with a possible increasing of the OD-pairs covered. However, the travel time of the OD-pairs initially covered could be increased due to the additional dwell time at the new station, and some of such pairs could not be covered by the modified network. That is, a new station leads to an opposite effect since simultaneously the objective value increases with new pairs covered, and decreases with the pairs which are lost. In the second place, a new station penalizes the travel time of those trips which maintain their combined plane-network mode with the new station, that is, those trips already covered by the initial network, and which continue covered after adding the new station. We say *permanent* trips to such trips.

For avoiding an excessive penalization on the travel time of permanent trips, we introduce a constraint on the increasing of their traveling time. With these considerations we formulate a trip covering location problem on a tree network, and propose a solution approach based on decomposing the problem in a collection of subproblems and, for each of them, identifying a subquadratic in the number of OD-pairs Finite Dominating Set.

# On location-allocation problems for dimensional facilities[*]

Lina Mallozzi,[1] Justo Puerto,[2] and Moisés Rodríguez-Madrena[2]

[1] *University of Naples Federico II, Naples, Italy,*   lina.mallozzi@unina.it

[2] *IMUS, Universidad de Sevilla, Sevilla, Spain,*   puerto@us.es
madrena@us.es

This work deals with a bilevel approach of the location-allocation problem ( [2]) with dimensional facilities. We present a general model that allows us to consider very general shapes of domains for the dimensional facilities.

## 1.     The problem

We are given a number of dimensional facilities that are shaped like general domains and the customers are distributed in the demand region according to a demand density that is an absolutely continuous probability measure. The goal is to locate them in a general planar demand region.

For a feasible location of the facilities, a partition of the demand region that determines the allocation of the customers to the facilities has to be done minimizing the total social cost. See e.g. [1] for similar partitions in a different context.

We are interesting in finding the best location of the facilities in such a way that the summation of certain realistic costs over all the facilities must be the cheapest possible, knowing that the partition of the customers is done as it is explained above. Among that costs, the congestion cost is computed once the partition of the customers in the demand region is done. These assumptions impose to our problem a hierarchical structure of bilevel problem.

---

[*]This research has been partially supported by Spanish Ministry of Economia and Competitividad/FEDER grants number MTM2016-74983-C02-01.

Different particular applications, which are discussed in our work, fit within this general problem. We prove the existence of optimal solutions for the bilevel problem under mild, natural assumptions. To achieve these results we borrow tools from optimal transport mass theory that allow us to give an explicit solution structure of the considered lower level problem.

## 2. Solution approaches

We propose a discretization approach that can approximate, up to any degree of accuracy, the optimal solution of the original problem. This discrete approximation can be optimally solved via a mixed-integer linear program. To address very large instance sizes we also provide a GRASP heuristic that performs rather well according to our experimental results. The graphical output of our algorithms for some illustrative test examples can be seen in Figure 1.

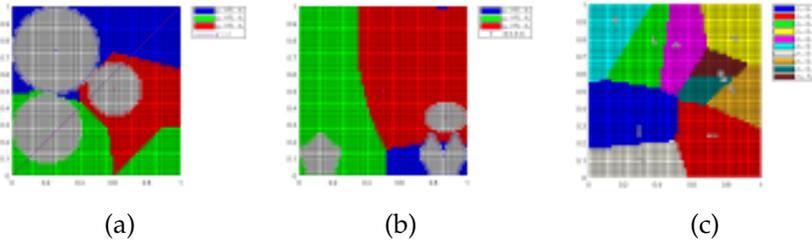

| (a) | (b) | (c) |

*Figure 1.*   Graphical output of the proposed algorithms for three different test examples.

# Robust feasible rail timetable

Ángel Marín,[1] Miguel A. Ruiz-Sánchez[1], Esteve Codina[2]

[1]*Instituto Matemática Interdisciplinar, Universidad Complutense Madrid, Spain*

[2]*Universitat Politécnica de Catalunya, Barcelona, Spain*

The railway timetable problem consists in selecting the optimal train routes and schedules to minimize the rail traffic service time. The growth in demand encourages rail managers to improve the effective use of the infrastructure occupations, but keeping the passenger service quality (i.e., travel time). Upgrading the infrastructure helps to achieve these objectives but increasing resource investment. The design of effective timetables may also help.

The timetable planning problem may be studied by means of two approaches: micro and macro. The macro approach simplifies the representation of the railway infrastructure: it considers railway segments linking consecutive control areas (stations, junctions, et.) as a single network node. This approach is suitable for tackling problems from the passengers' point of view. The micro approach makes use of a detailed description of the railway control infrastructure and considers the potential traffic conflicts in the control area.

## 1.    Macro Timetable

The macro railway timetable will be defined under the passengers or operative point of view. A set of platforms joint with the entrée and leave tracks is considered a station, and a set of multiple tracks crossing in a node is a junction. Both will be considered in the micro model as control zones. The macro model tries to minimize the optimal train travel time, considering the passenger travel time and the delays at destination. The macro does not consider the conflicts at the control zones, so the solution will be not feasible considering those.

A timetable to be feasible needs to be conflict-free: trains must be able to travel at their planned velocity without ever having to stop or slow down due to restrictive signals. To ensure feasibility timetable, the micro model allow longer stays at platforms or special tracks, where the trains have planned stops.

## 2. Micro Timetable

The micro model is defined in terms of track-circuits, they are track segments on which the presence of a train is automatically detected. Sequences of track-circuits are grouped into block sections, which access is controlled by signals.

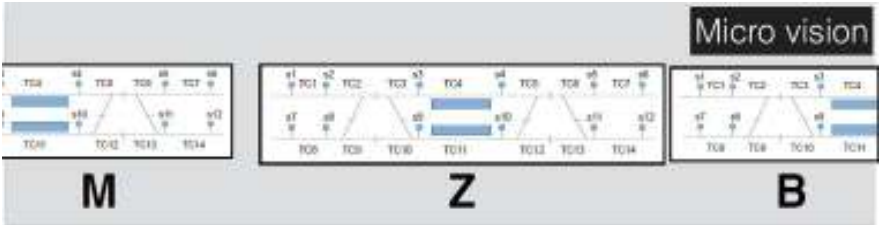

*Figure 1.* Micro vision services Madrid-Barcelona.

We define the routes in control areas in terms of a sequence of track-circuits and by intermediate stops. A sequence of track-circuits can be transversed performing or not intermediate stops defining different routes. The objective function of the micro models in each control area is to minimize the shifted time between the arrival and departure times given by the macro optimization model, which need to obtain micro-feasible solutions. In short, the micro constrains may be described by: the train occupation, reservation and stay management, plus the train routing and scheduling in the control areas.

An integrated model may tackle both approaches at once. The macro model plan the arrival and departure timetable using a tentative train time, providing it at micro level. The micro model makes feasible and robust this timetable by computing the optimal routing and schedule of the train at the control areas, minimizing the shifted arrival and departure at each of them. The essential micro inputs are the arrival and departing of each train at each control area. The micro outputs are the shifted arrival and depart of each train at each control area, so that the train route and sequence free-conflicts in the area.

The essential macro inputs are the shifted arrival and departure train times at the control area. The macro outputs are the arrival and departure of the trains, joint with the generalized costs perceived by the passengers and the scheduled delay of the trains being early and lateness at the control areas.

# Wildfire Location Model: A new proposal

Juan A. Mesa,[1] Mariano Marcos-Pérez.[1]

[1]*Escuela Técnica Superior de Ingenieros. Departamento de Matemática Aplicada II, Universidad de Sevilla, Camino de los Descubrimientos s/n, 41092 Seville, Spain*

Nowadays there are an increasing awareness of environmental problems. Wildfires pose a serious threat to communities and ecosystems throughout the world. Wildfires containment and locations of limited resources to mitigate the impact of natural disasters are important but challenging tasks. The location of a set of resources when fighting fire is proposed it is one of the many problems we need to improve in order to get a better solution or at least more useful. Wildfire suppression combines multiple objectives, the main objective of forest-fire management is to minimize the damage caused by forest fires. Empirical studies have identified several factors that affect the development of a forest fire. The prevailing meteorological conditions of the area under consideration, that evidently play an important role in the location of firefighting resources, have been completely ignored or inadequately considered.

There are many models which can be improved adding a wind restriction. Karkazis (1992) [1] presented the most recent wind-discrete models and solution methods to locate facilities causing airbone pollution. Hodgson and Newstead (1978) [2] developed two location-allocation models for assigning a limited number of airtanker. Dimopoulou and Giannikos (2001) [3], using information provided by a GIS have determined the optimal location of fire-fighting resources. Belval et al. (2015) [4] have presented a mixed integer program to model spatial wildfire behaviour and suppression placement decisions. Alvelos (2018) [5] presents the location of a set of resources when fighting fire is proposed.

In this work, we present a wildfire location model including a wind-discrete model. This model is intended to be a future supplement in resource planning to get a better results when we fight against fires. The purpose of investigations such as this is to improve forest science using Operation Research and Optimization, the future of the forest science is largely a question of cooperation between different branches of study, in this case, mathematics.

# Optimal allocation of fleet frequency for "skip-stop" strategies in transport networks

Juan A. Mesa,[1] Francisco A. Ortega,[2] Ramón Piedra-de-la-Cuadra[3] and Miguel A. Pozo[4]

[1] *Universidad de Sevilla, Sevilla, Spain,* jmesa@us.es

[2] *Universidad de Sevilla, Sevilla, Spain,* riejos@us.es

[3] *Universidad de Sevilla, Sevilla, Spain,* rpiedra@us.es

[4] *Universidad de Sevilla, Sevilla, Spain,* miguelpozo@us.es

The planning of the public transport systems includes the design of the lines and the frequency of the services of transport. The occasional incidents in the functioning of the system generally are not considered in the initial planning. To reduce the disturbing effect, the operator of a service of passengers' transport must be able to implement some strategy of control to fit the schedules to the conditions of the real time traffic. The strategies more commonly used are the express service (certain stations of the corridor skip due to little passengers' flow), short-turning, or deadheading and the combination of different actions of control [3].

In this research work a methodology is developed to implement a redistribution of services along a line of railway traffic, which must be carried out by the operator choosing new train schedules within a series of feasible space-time windows, previously established by the railway infrastructure manager. The objective is to minimize the loss of users, who could perceive a worsening in the quality of the service that until now they had been receiving. In addition, the result obtained in that first phase will condition the subsequent decision of the transport operator, consisting of classifying the fleet of its trains into two types (A and B) and the stations, at the same time, in three types of stops: type A, where will stop the trains identified with the distinction A; type B, where the trains identified as mode B will stop; Type AB, where both kinds of trains will stop. This double allocation, both for trains and for stations, is called the Skip-Stop strategy.

# 1.    Skip-Stop Strategy and Knapsack Problem

The Skip-Stop mechanism consists of privileging a larger number of passengers offering shorter travel times, as a result of having previously selected a group of low-activity stations, where trains wouldn't stop to pick up or let off passengers

The travel time between stations in a railway line consists of five components, identified as phases of acceleration, constant speed, inertia, braking and downtime, in [2] can find the algebraic expressions habitually used to calculate these times. In consequence, the operation of omitting stops reduces the travel time for the users within the vehicle and increases the speed of operation in the provision of the service. However, some users will experience a longer time of waiting, accessing, exiting and, possibly, transferring. Therefore, there is no guarantee that any skipping operation will decrease the total travel time of the potential travellers. The selection and coordination of stops must be made by using a criterion according to an objective function.

We propose, in this work, to model an existing problem (Skip-Stop problem) through the Knapsack Problem (KP) taking advantage of the large amount of material available from the KP. The solution of the Skip-Stop problem will consist of two phases; in the first, we find the optimal strategy of skipping stops for a given train fleet and, in the second phase, we determine, by means of a heuristic, the optimal allocation for train types. For this last purpose, we will develop the concept of proximity between the railway routes and, in accordance with the Hall method [1], design a Matheuristic that optimizes the Skip-Stop Strategy.

# Introduction to planar location with orloca

Manuel Munoz-Marquez,[1]

[1]*Cadiz University, Cadiz, Spain*  manuel.munoz@uca.es

The `RcmdrPlugin.orloca` package devoted to solve the planar continuous location problem is presented. It has been developed for `R` as free software.

It is intended to be used as an easy way to introduce the planar point location problems to the students. This is done providing a GUI and an on-line interactive application to handle and solve such problem.

## 1.     Introduction

In a location problem, we seek the optimal location of a service. Examples of localization problems are: find the optimal location of the central warehouse or an ambulance that must attend to the patients.

The package solves the problem of locating a single point in the plane, minimizing of the sum of the weighted distances to the demand points. New versions of the package will include new location models. The package `RcmdrPlugin.UCA` [1] provides a GUI to do that.

## 2.     Package features

A new class of objects, designated as `loca.p`, has been defined to handle instances of the problem.

From the menu one can create new instances of `loca.p` objects, generate new random instances and make summaries of the data. One can also evaluate the weighted average distance, in the following `distsum` and calculate the gradient of it.

There is also an option to find the minimum of `distsum` using the Weiszfeld algorithm, see [2] or [3] , or a global optimization one. The Weiszfeld algorithm includes a test for optimality for demand points.

Four options has been provided to make some plots as the figures show.

The help menu provides several options to get more information about the use of the package and provides several examples.

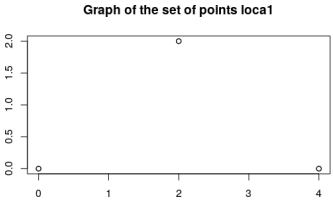

*Figure 1.*   Demand point set

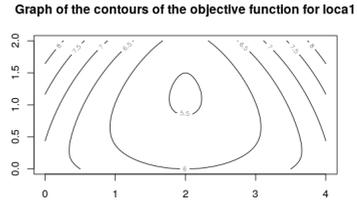

*Figure 2.*   Contour plot

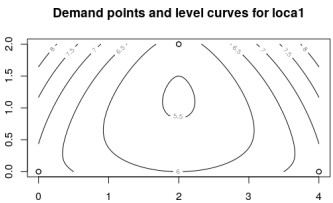

*Figure 3.*   Demand and contour plot

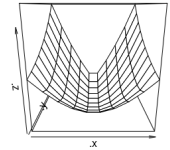

*Figure 4.*   3D plot

## 3.   Conclusions

The RcmdrPlugin.orloca package solves the planar location problem of a single service using an user friendly menus, so it is a good way to introduce this problem to students. More info in http://knuth.uca.es/orloca and an on-line interactive demo in http://knuth.uca.es/shiny/orloca/.

# Emergency Vehicle Location Model considering uncertainty and the hierarchical structure of the resources

José Nelas,[1] Joana Dias[1,2]

[1]*Faculdade de Economia, Universidade de Coimbra, Portugal*

[2]*CeBER and Inesc-Coimbra, Universidade de Coimbra, Portugal,*   joana@fe.uc.pt

The main goal of emergency services is to guarantee that help arrives to populations where and when it is needed. Depending on the severity of the emergency episodes, it is possible to define maximum time limits that should be respected to assure a proper and timely assistance. The location of emergency vehicles is crucial for guaranteeing that this goal is achieved. In this work we present a mathematical model that considers the location of emergency vehicles under uncertainty. Moreover, the model takes explicitly into account the hierarchical features of these vehicles, namely considering that some types of vehicles can substitute others that are not available when the episode occurs.

## 1.     Introduction

The location of emergency resources has been a subject studied by several different researchers in the last decades. One of the most used type of model is the covering location model: the objective is to guarantee that all the population is covered, considering a given distance or time limit between each population and the nearest available resource. When full coverage is not possible, due to budget restrictions, for instance, then one of the approaches is to consider the maximization of the covered population. This choice can have as consequence severe inequalities, especially considering the most distance and depopulated regions. While the initial models were deterministic and static, recent developments have recognized the importance of considering the inherently stochastic nature of the problem, as well as time. When dealing with emergency resources, it is very important to acknowledge that not all the resources will be available at all times, namely because they can already be assigned to an occurrence.

# 2.      The Proposed Model

In the developed model, uncertainty is represented through a set of scenarios. Each scenario is characterized by having a given set of different emergency episodes, that occur in different locations, at different time periods, and that have different levels of severity (thus requiring different types and number of vehicles).

It is necessary to decide where to locate the vehicles (this decision is not scenario dependent), and the way in which vehicles are assigned to emergency occurrences (these decisions are scenario dependent). An occurrence is considered "covered" if it is possible to assign to it all the required vehicles, within the defined time limits. A coverage matrix determines, for each level of assistance, which vehicles are within the established time limit from the emergency occurrences. The objective function considers the maximization of the covered occurrences.

Time is considered in the model through the use of an incompatibility matrix that, for each scenario, each pair of emergency occurrences and each type of vehicle determines whether these occurrences have or have not overlapping time periods. If there is no overlapping time periods, then the same vehicle can be assigned to both occurrences. If there is at least one overlapping time period, this is not possible anymore (if a vehicle is assigned to one of the occurrences it cannot be assigned to the other). There is also information about the possibility of a given type of vehicle being able to substitute another one, if necessary. A vehicle capable of providing more differentiated emergency care can also be assigned to an occurrence that would only need a less differentiated care, for instance. When deciding the assignment of vehicles to occurrences, this flexibility is also considered.

The developed mathematical model will be presented, as well as some illustrative examples and preliminary computational results.

# Solving the Ordered Median Tree of Hubs Location Problem

Miguel A. Pozo,[1] Justo Puerto,[2] and Antonio M. Rodríguez-Chía,[3]

[1]*Universidad de Sevilla, Spain.*,   miguelpozo@us.es

[2]*Universidad de Sevilla, Spain.*,   puerto@us.es

[3]*Universidad de Cádiz, Spain.*,   antonio.rodriguezchia@uca.es

Standard Hub Locations Problems assume that inter-hub connections between an origin-destination pair can be routed through one or at most two hubs. However, it has been observed by several authors that in many applications the backbone network is not fully interconnected [1,2] or it can even be not necessarily connected.It is of special interest the case where the underlying interconnection network is connected by means of a tree. Such problem is called the Tree of Hubs Location Problem and was introduced by Contreras et al ( [3,4]).

Another feature, namely weighted averaging objective functions, has also been incorporated to the analysis of Hub Locations Problems [5,6]. It has been recognized as a powerful tool from a modeling point of view because its use allows to distinguish the roles played by the different entities participating in a hub-and-spoke network inducing new type of distribution patterns. Each one of the components of any origin-destination delivery path gives rise to a cost that is weighted by different compensation factors depending on the role of the entity that supports the cost. This adds a "sorting"-problem to the underlying hub location problem. The objective is to minimize the total transportation cost of the flows between each origin-destination pair after applying rank dependent compensation factors on the transportation costs.

In this paper, we study the Ordered Median Tree of Hub Location Problem (OMTHLP). The OMTHLP is a single allocation hub location problem where $p$ hubs must be placed on a network and connected by a non-directed tree. Each non-hub node is assigned to a single hub and all the flow between origin-destination pairs must circulate using the links connecting the hubs. The objective is to minimize the sum of the ordered weighted averaged assignment costs plus the sum of the circulating flow costs. We will present different MILP mathematical formulations for the OMTHLP based on the properties of the Minimum Spanning Tree Problem and the Ordered Median optimization. We establish theoretical and empirical comparisons between these new formulations and we also provide reinforcements that together with a proper formulation are able to solve medium size instances on general graphs.

# Feasible solutions for the Distance Constrained Close-Enough Arc Routing Problem

Miguel Reula [*1], Ángel Corberán[1], Isaac Plana[2] and José Maria Sanchis [3]

[1]*Dept. d'Estadística i Investigació Operativa, Universitat de València, Spain*

[2]*Dept. de Matemáticas para la Economía y la Empresa, Universitat de València, Spain*

[3] *Dept. de Matemática Aplicada, Universidad Politécnica de Valencia, Spain*

The Close-Enough Arc Routing Problem (CEARP), also known as Generalized Directed Rural Postman Problem, is an arc routing problem with interesting real-life applications, such as routing for meter reading. In this application, a vehicle with a receiver travels through a series of neighborhoods. If the vehicle gets within a certain distance of a meter, the receiver is able to record the gas, water, or electricity consumption. Therefore, the vehicle does not need to traverse every street, but only a few, in order to be close enough to each meter. We deal with an extension of this problem, the Distance-Constrained Close Enough ARP, in which a fleet of vehicles with distance constraints is available. The vehicles have to leave from and return to the depot, and the length of their routes must not exceed a maximum distance (or time). Several formulations and exact algorithms for this problem were proposed in [1]. Since the size of the instances solved to optimality is far from those arising in real-life problems, we propose here a multi-start heuristic algorithm with an improvement phase that incorporates an effective exact procedure to optimize the routes obtained. In order to assess the relative efficiency of our algorithm, extensive computational experiments have been carried out. The results show the good performance of the proposed heuristic, even in the instances with a very tight maximum distance.

* miguel.reula@uv.es

# Steiner Traveling Salesman Problems: when not all vertices have demand

Jessica Rodríguez-Pereira,[1,2] Enrique Benavent,[3] Elena Fernández,[2] Gilbert Laporte,[1] and Antonio Martínez-Sykora[4]

[1]*Canada Research Chair in Distribution Management, HEC Montréal, Canada*

[2]*Department of Statistics and Operation Research, Universitat Politècnica de Catalunya-BcnTech, Spain*

[3]*Department of Statistics and Operation Research, Universitat de Valencia, Spain*

[4]*Southampton Business School, University of Southampton, United Kingdom*

The purpose of this work is to present a new compact formulation and efficient exact solution algorithm for the Steiner Traveling Salesman Problem (STSP) on an undirected network and its location extension. The STSP is an uncapacitated node-routing problem looking for a minimum-cost route that visits a known set of customers with service demand, placed at vertices of a given network, which is assumed to be uncomplete. Although not all the vertices in the network have demand, some non-demand vertices may have to be visited for connecting demand vertices served consecutively in a route. Thus, the specific set of vertices that must be traversed in feasible routes is not known in advance. The location extension, LSTSP, studies the case when several depots are allowed and their location has to be decided as well.

We propose compact mixed integer linear programming formulations for the STSP and the SLTSP. All formulations are defined on the original undirected graph and use a small number of two-index decision variables only. They exploit the property that there is an optimal solution, both for the STSP and the LSTSP, where no edge is traversed more than twice, and use two sets of binary variables only, which are associated with the first and second traversal of edges, respectively. Feasibility of solutions is modeled with two families of constraints of exponential size, one for the connectiv-ity with the depot and another one for the parity of the visited vertices. For the parity of the vertices we use an adaptation of co-circuit constraints, which exploits the relationship between our two sets of decision variables. While co-circuit inequalities are nowadays very often used to model parity in arc routing problems, we are not aware of any node-routing problem where such inequalities have been used to model the parity of visited vertices. Since the two-index decision variables do not associate traversals of edges with the facilities of the routes they belong to, the LSTSP formulation requires an additional set of constraints that involve the location decision variables as well, in order to guarantee that routes are well defined and return to their starting location.

We have developed an exact branch-and-cut algorithm for the STSP that allows us to optimally solve instances with up to 500 vertices in very moderate computing times. Our computing times never exceed 350 seconds for instances with up to 250 vertices and, except for two out of 60 instances, do not exceed our time limit of 7200 seconds for the larger instances with a number of vertices ranging in 275-500.

We have also developed an efficient exact branch-and-cut algorithm to solve the two-index formulation for the SLTSP. As could be expected, the computational effort required to solve the instances is now considerably larger than for the STSP. Still, SLTSP instances with up to 500 vertices and 10 potential locations were optimally solve within the maximum allowed computing times 7200 seconds.

# Addressing locational complexity: network design and network rationalisation

Diego Ruiz-Hernandez,[1] Jesus M. Pinar-Pérez,[2]
and Mozart B.C. Menezes[3]

[1]*Operations Management and Decision Sciences Division, Sheffield University Management School, S10 1FL, Sheffield, UK*  d.ruiz-hernandez@sheffield.ac.uk

[2]*Dept. of Quantitative Methods, University College for Financial Studies, 28040, Madrid, Spain*  jesusmaria.pinar@cunef.edu

[3]*Operations Management and Information Systems Dep., Kedge Business School, 33405, Bordeaux, France*  mozart.menezes@kedgebs.com

Facility location problems are well known combinatorial problems where the objective is to minimize certain measure of the cost incurred for (or the benefit attained from) serving customers from a set of facilities. A typical location problem will either aim at maximising the demand covered by strategically locating a given number of facilities, or at finding the optimal number and location of facilities necessary for satisfying the total demand in a region.

Our aim is to bring to the field of facility location the concept of structural complexity, opening up a new research line. Broadly speaking, structural complexity refers to the negative effects of the proliferation of products, distribution channels and markets. Focusing on locational complexity, the main objective of this project is to create awareness about the need of considering complexity issues –and their impact on profitability- when deciding the location and size of a distribution network. The rationale behind our argument is that an oversized distribution network may cause hidden costs that hinder the capacity of the supply chain for translating revenue into bottom-line benefits.

In this work, using an entropy based measure for structural complexity developed by the authors in previous research [1], we propose an variant of the traditional p-median problem that includes a complexity parameter in the model's formulation, the K-MedianPlex problem:

$$\max_{S \subset N : |S| = K} \quad Z_{Plex}^K = \sum_{k \in S} R^{(k)} \left(1 - \alpha C_p^{(k)}\right) - \phi K$$

with

$$
\begin{aligned}
C_p^{(k)} &= \sum_{i \in \mathcal{N}_k} w_i \log_2 \left(\frac{1}{w_i}\right), \quad k \in S \\
R^{(k)} &= \sum_{i \in \mathcal{N}_k} \left(r - \gamma d_{ik}\right) W_i, \quad k \in S \\
\alpha &: \quad \alpha C_p^{(k)} < 1, \quad k \in S
\end{aligned}
$$

where $\mathcal{N}$ is the set of network nodes; $S \subset \mathcal{N}$, the set of open facilities; $W_i$, the weight of demand node $i \in \mathcal{N}$; $\phi$, a fix facility cost; $r$, the revenue per unit; $\alpha$, a profit loss factor due to complexity; $\gamma$, a generic transportation cost; and $w_i = W_i / \sum_{i \in \mathcal{N}} W_i$ for all $i \in \mathcal{N}$.

Given the strongly combinatorial nature and non-convexity of the objective function, we propose an algorithmic approach based on solving a K-median problem and sequentially reasigning demand nodes across facilities and solving local 1-Median problems aiming at maximising the function $Z^{Plex} \left(\mathcal{N}^{|S'|}, S'\right)$, where $\mathcal{N}^{|S'|}$ represents the collection of allocation sets associated to a given solution $S'$.

However, location complexity is not typically a result of network design, but a problem that arises from successive network expansions aimed at capturing market share. In order to reduce complexity, firms may find it profitable abandoning certain markets (although they are usally reluctant under the rationale that lost sales will affect profit negatively). We propose a strategy for succesively uncovering demand nodes until no profit improvement can be further attained.

Experimental results suggest that higher profits can be attained by reallocating demand nodes across facilities, relocating facilities and/or eliminating non-profitable demand nodes. As it may be expected, the improvement routines return better results for high values of the complexity cost parameter $\alpha$, and for larger transportation costs.

# Using a kernel search heuristic to solve a sequential competitive location problem in a discrete space [*]

Dolores R. Santos-Peñate,[1] Clara M. Campos-Rodríguez,[2] and José A. Moreno-Pérez[3]

[1]*Instituto de Turismo y Desarrollo Económico Sostenible/Dpto de Métodos Cuantitativos en Economía y Gestión, Universidad de Las Palmas de G.C., 35017 Las Palmas de Gran Canaria. Spain,* dr.santos@ulpgc.es

[2]*Instituto Universitario de Desarrollo Regional. Universidad de La Laguna. 38271 La Laguna. Spain,* ccampos@ull.es

[3]*Instituto Universitario de Desarrollo Regional. Universidad de La Laguna. 38271 La Laguna. Spain,* jamoreno@ull.es

In the leader-follower, $(r|p)$-centroid or Stackelberg location problem, two players sequentially enter the market and compete to provide goods or services. This work considers this competitive facility location problem in a discrete space. The customer choice rule is defined as a possibility function. For each customer, an S-shaped function is used to share the demand among competitors when the difference in distances to the competing firms is small. To solve the problem, the linear programming formulation for the leader and the follower are integrated into an algorithm which, in an iterative process, finds a solution by solving a sequence of these linear problems. We propose a matheuristic procedure that provides solutions for the leader via a kernel search algorithm. The proposed solution approach is illustrated with some computational results obtained for the binary rule and different S-shaped customer choice functions. The results obtained with an exact algorithm and the heuristic procedure are compared.

---

[*]This study was partially funded by Ministerio de Economía y Competitividad (Spanish Government) with FEDER funds, through grants ECO2014-59067-P and TIN2015-70226-R, and also by Fundación Cajacanarias (grant 2016TUR19).

# Author Index

# IX Work
# Locat
# Analysis
# Proble

SPONSORS:

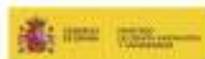

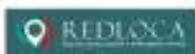

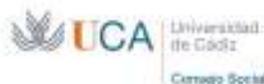

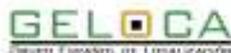

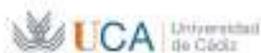